 \documentclass[preprint2]{aastex}

\slugcomment{Astronomical Journal, submitted}

\shorttitle{UV Attenuation in Galaxies}
\shortauthors{Keel et al.}

\begin{document}


\title{The Ultraviolet Attenuation Law in Backlit Spiral Galaxies\footnote{Based in part on observations made with the NASA Galaxy Evolution Explorer. 
GALEX is operated for NASA by the California Institute of Technology under NASA contract NAS5-98034.}}


\author{William C. Keel \altaffilmark{1,2}}
\affil{Department of Physics and Astronomy, University of Alabama, Box 870324, Tuscaloosa, AL 35487}
\email{wkeel@ua.edu; Twitter @NGC3314}

\author{Anna M. Manning\altaffilmark{1}}
\affil{Stennis Space Center}

\author{Benne W. Holwerda}
\affil{ESA-ESTEC, Keplerlaan 1, 2201 AZ Noordwijk, The Netherlands}
\email{bholwerd@rssd.esa.int; @BenneHolwerda}

\author{Chris J. Lintott\altaffilmark{1}}
\affil{Astrophysics, Oxford University; and Adler Planetarium, 1300 S. Lakeshore Drive, Chicago, IL 60605}
\email{cjl@astro.ox.ac.uk; @chrislintott}


\and

\author{Kevin Schawinski}
\affil{Institute for Astronomy, ETH Z\"urich,
Wolfgang-Pauli-Strasse 27,
CH-8093 Zurich,
Switzerland}
\email{kevin.schawinski@phys.ethz.ch; @kevinschawinski}


\altaffiltext{1}{Visiting astronomer, Kitt Peak National Observatory, National Optical Astronomy Observatories, which is operated by the 
Association of Universities for Research in Astronomy, Inc. (AURA) under cooperative agreement with the National Science 
Foundation. The WIYN Observatory is a joint facility of the University of Wisconsin-Madison, Indiana University, Yale University, 
and the National Optical Astronomy Observatory.}
\altaffiltext{2}{SARA Observatory}

\newpage

\begin{abstract}
The effective extinction law (attenuation behavior) in galaxies in the emitted ultraviolet regime is well known
only for actively star-forming objects and combines effects of the grain properties, fine structure
in the dust distribution, and relative distributions of stars and dust. We use GALEX, XMM Optical Monitor, and HST data
to explore the UV attenuation in the outer parts of spiral disks which are backlit
by other UV-bright galaxies, starting with the candidate list of pairs provided
by Galaxy Zoo participants. New optical images help to constrain the
geometry and structure of the target galaxies. Our analysis incorporates galaxy symmetry, 
using non-overlapping regions of each galaxy to derive error estimates on the attenuation
measurements. The entire sample has an attenuation law across the optical and UV which
is close to the \cite{calzetti} form; the UV slope for the overall sample is substantially shallower than found by \cite{Wild},
which is a reasonable match to the more distant galaxies in our sample but not to the weighted combination including NGC 2207. 
The nearby, bright spiral NGC 2207 alone gives accuracy almost equal to the rest of our sample, and its outer arms 
have a very low level of foreground starlight. Thus, this widespread, fairly ``grey" law
can be produced from the distribution of dust alone, without a necessary contribution from differential escape of
stars from dense clouds. Our results indicate that the extrapolation needed to compare attenution between
backlit galaxies at moderate redshifts from HST data, and local systems from SDSS and similar data, is mild
enough to allow use of galaxy overlaps to trace the cosmic history of dust in galaxies. For NGC 2207,
HST data in the near-UV F336W band show that the covering factor of clouds with small optical attenuation
becomes a dominant factor farther into the ultraviolet, which opens the possibility that widespread diffuse dust dominates
over dust in star-forming regions deep into the ultraviolet. Comparison with published radiative-transfer models indicates that the
role of dust clumping dominates over differences in grain populations, at this coarse spatial resolution.
\end{abstract}




\section{Introduction}
Our understanding of the effects of dust grains in galaxies has increased dramatically with such new capabilities as
sensitive far-infrared measures, spatially resolved modeling of the spectral-energy distributions (SEDs) of star/grain
mixes, and photometry of resolved galaxies giving independent reddening maps from the stars themselves (\citealt{Berry2012},
\citealt{PHAT}). The emerging
starlight is modified in ways which depend crucially on the relative distributions of stars and dust, and on the small-scale
structure in the dust.

These factors produce proportionally greater uncertainties in the emitted ultraviolet range, affecting much of the data
relevant to galaxy evolution. As extensive HST surveys are allowing exploration of the
evolution of galaxy morphology, they can also help track the evolution of the dust content of galaxies if we can
connect UV and optical attenuation measures. Existing models have a wide range in the predicted behavior of both
dust mass and resulting attenuation with redshift \citep{CalzettiHeckman1999}, since dust production, destruction, and
the shrinking ISM mass fraction compete in ways that are not well constrained at large redshifts. Broadly, attenuation
from SED fits and photometric redshifts has shown a peak at $z \approx 1.5$, declining at earlier and later epochs \citep{RR2003}.
Observationally, fits to SEDs from the UV to FIR by 
\cite{elais} indicate that the dust content of low-mass galaxies has increased from $z=0.7$ to the present,  while galaxies
at high stellar mass show no such trend. Deep surveys over wide spectral ranges are now deep and wide enough
to test models for the evolution of dust. 
Connection of the history of dust mass to observables must fold in at least implicit knowledge of the dust distribution on both large and small 
scales.

A result which has found wide applicability is the effective extinction law derived by \cite{calzetti},
based on the SEDs of star-forming galaxies. It is
relatively flat with wavelength as compared to the behavior found from
star-by-star studies, implying that it is strongly affected by
the relative distributions of stars and dust and by unresolved fine structure in the
dust distribution itself. The UV range is particularly sensitive to these effects, due to
the short lifetimes of the stars which dominate the UV light from star-forming systems, and
to the inevitable bias in favor of more transparent areas within a finite region of a galaxy
\citep{Fischera}. \cite{KW2001a} found that that measured reddening behavior in two backlit spiral
galaxies becomes flatter (grayer) when the data are smoothed over successively larger regions before analysis.
Effects of dust structure must be included in SED models in order to retrieve either the intrinsic stellar SED
or the effective extinction. Comparison of galaxy disks seen at various angles can provide independent
information on some of the distribution issues (i.e. \citealt{Wild}). 

This issue highlights the distinction between the observable quantity attenuation,
measured on some size scale,  and the actual
extinction attributable to the grain properties, which is typically manifested in what are essentially
point-source measurements of light from individual stars, where mixture and scattering effects are
negligible \citep{WittGordon2000}. In interpreting observations of galaxies, the net loss of direct starlight is often termed extinction,
although attenuation or effective extinction are more precise description. 
Absorption is a property of the grains, while extinction
includes both actual absorption and scattering out of the line of sight; we actually measure attenuation (sometimes 
described as effective extinction), combining these with
scattering into the line of sight, potentially important for large areas within galaxies. These essentially coincide when
deriving extinction from individual stars, since scattering from a diffuse medium can be negligibly small compared to
the light of a star.

For large regions of a galaxy, scattering may be important. Scattering within
each galaxy of a pair will be removed by symmetry, so that the asymmetric component of scattering of light from the background
galaxy into the line of sight by grains in the foreground system remains a potential concern. As shown by
\cite{WKC2000}, this component drops very rapidly with galaxy separation; it also drops with increased clumping of the dust \citep{WittGordon2000}. 
The strong forward scattering
of local grain populations implies that the scattering contribution may drop strongly again at small
galaxy separations as the required scattering angle grows. We note that none of the optical images
of our targets show the characteristic bluing toward the edges of dust lanes which is a sign of scattering

We have explored the use of backlit galaxies to measure dust attenuation for
local systems, with the utility of the technique extended enormously with
the production of a catalog of nearly 2000 such pairs \citep{Keel2013} based on examination of 
Sloan Digital Sky Survey (SDSS) images by volunteers within the Galaxy Zoo project \citep{gz2008}.
This large starting sample allows us to select significant subsamples based
on galaxy type or geometry. For this study, we concentrate on systems that are otherwise
of limited value in dust studies - spiral/spiral pairs, where the high UV flux of the background spiral
makes up for its lack of detailed symmetry compared to E/S0 background systems.
Errors in estimating the light loss are unavoidably larger than in the optical using E/S0 background
galaxies, so we combine data from multiple systems and, where possible, average along
the spiral pattern to suppress fluctuations due to background structure. A key
feature of this backlighting approach is that it is weighted by area
rather than by the luminosity of embedded stars, values, which is clearly more
representative for background sources and may be most directly representative of 
the effect in emerging radiation when averaging over regions of a galaxy several kpc in size.
This approach is explicitly sensitive to a grey component, which is left poorly
constrained by purely spectral techniques, and is sensitive to arbitrarily cold dust components.
These features make it a useful complement to, for example, far-IR and submillimeter survey results.

For most of our studies of dust in backlit galaxies, we have fairly strict symmetry requirements,
so we can trace radial behavior. For this project, since we are most interested in the wavelength
behavior of attenuation, we can relax this requirement as long as we can quantify the effects of
galaxy asymmetry on our derived attenuation values. Thus, our sample here includes some galaxies
showing mild effects of interactions (most notably NGC 2207/IC 2163).

In this paper, we report an extension of backlighting measurements into the ultraviolet, where most previous attenuation
results within galaxies are limited either to very nearby systems (\citealt{Bianchi}, \citealt{PHAT}) or to actively star-forming galaxies
\citep{calzetti}. One motivation for this study is the promise of using similar techniques in deep HST images to compare
the dust signatures across a a significant redshift range, where the typical scaling with wavelength might become a dominant
factor in comparison with the nearby Universe. Our results trace the UV attenuation into the outer regions of disks, in some cases to
areas where the foreground surface brightness is so low that the properties are not affected
by features of the foreground galaxies. 
Where absolute quantities are important, we use the WMAP ``consensus cosmology" \citep{WMAP},
values, notably H$_0 = 72$ km s$^{-1}$ Mpc$^{-1}$.

\section{Observations}
\subsection {Galaxy Sample}
To measure dust effects using the backlighting approach in the ultraviolet , we need background galaxies that are UV-bright, which in practice means 
spiral systems (with their attendant lack of exact symmetry). We considered spiral/spiral galaxy pairs from the 
large catalog of overlapping-galaxy pairs generated from
Galaxy Zoo candidates \citep{Keel2013}. A subset was inspected in detail, based on availability of
long GALEX exposures (integration $>600 $ seconds in at least the NUV band), large 
angular size, and suitability for dust analysis based on symmetry and the geometry of overlap, 
as evaluated from higher-resolution optical images. Many of these candidates proved unsuitable
for dust analysis due to low UV surface brightness in the region of interest, lack of symmetry
in the UV, or unfortunate location of UV-bright star-forming regions (identifiable by color as well as
brightness).

\subsection {Data:}
\subsubsection {Ultraviolet}
Ultraviolet imagery came primarily from the GALEX archive \citep{GALEX}. GALEX
carried a 0.5m telescope,
using a dichroic beamsplitter to simultaneously observe in both NUV (1925-2730 \AA\  
at half-peak) and 
FUV (1410-1640 \AA) bands. The resolution is approximately 5.0$^{\prime \prime}$ FWHM.
For some targets, observed after failure of the far-UV detector, only NUV 
data are available; for others, the total NUV exposures are significantly longer than
for FUV. These factors, combined with the increasingly clumpy structure of galaxies toward shorter
wavelengths, means our useful sample size shrinks and error
bars grow corresponding between NUV and FUV bands. In practice, we found that only 
exposures 600 seconds and longer provided sufficient signal-to-noise for our
analysis.

For the nearby, bright pair NGC 2207/IC 2163, we use data from the XMM-Newton Optical/UV Monitor
(OM; \citealt{XMMOM}). Its 0.3m primary mirror focuses the image onto a microchannel
plate whose output is rapidly read via an intensified CCD, covering a field of 16.2$^\prime$ square
at 0.48$^{\prime \prime}$/pixel. Relevant UV filters for our program are UVW1 (2450-3200 \AA),
UVM2 (2050-2450 \AA), and U, with effective wavelength
near 3440 \AA\ . The PSF for the XMM-OM is tighter than
for GALEX, with FWHM $\approx 3$$^{\prime \prime}$ in the UV filters. Long integrations are available
in the NGC 2207 field: 13380 seconds in UWM2 (roughly corresponding to GALEX NUV) 
and 8920 seconds in UVW1 (somewhat longer in wavelength than GALEX NUV).
The XMM-OM data in $B$, and to a smaller extent $U$,
were affected by a reflection artifact from an out-of-field source superimposed the southern part of NGC 2207 \citep{XMMhandbook};
the $U$-band reflection is minor enough that we can analyze this image, but we use the ground-based $ug$ data
rather than XMM-OM $B$. Suitable data are not available to extend analysis of this system to shorter wavelengths; GALEX
observed NGC 2207 in the FUV only during its all-sky survey, and a long NUV exposure was obtained only
after failure of the FUV detector.

\subsubsection {Optical}
We have obtained new optical images of many candidate galaxy pairs; often the higher
resolution and signal-to-noise ratio help make the interpretation clear, even when
the attenuation measurement for this project is limited to the resolution of the UV images.
When no absorption is detected in the optical images, we infer that the galaxy in question
lies in the background of the pair. The new images came from several sources.

Most images came from the 3.5m WIYN telescope with OPTIC fast-guiding camera \citep{OPTIC}, which
provides a 10$^{ \prime}$ field sampled with 0.14$^{\prime \prime}$ pixels. We
used this system in three sessions from April 2008 to May 2010.
Image quality usually ranged from 0.5-0.8$^{\prime \prime}$ FWHM; for most fields, bright stars were available allowing us to use the system's on-chip high-speed guiding. Passbands were $B$ and $I$; fringing from
night-sky emission in the $I$ was corrected using median-combined frames to generate a reference
pattern. For most pairs, exposures were $2 \times 10$ min in $B$ and 10 min in $I$.

The KPNO 2.1m telescope was used at its $f/8$ Ritchey-Chretien focus, in April and May 2012. A $2048 \times 2048$
TI CCD provided a field of 11$^{ \prime}$ at 0.305$^{\prime \prime}$/pixel, slightly vignetted at one edge.
Galaxy pairs were generally observed in $B$ (20 min) and $R$ (10 min) bands. 

For the nearby southern pair NGC 2207/IC 2163, we used the SARA-S remotely-operated 0.6m 
telescope at Cerro Tololo.
An Apogee camera with a $1024 \times 1024$ Kodak CCD provided a field 10.3$^{\prime}$ on a side at 0.61$^{\prime \prime}$/pixel, with $ugriz$ filters.
Image quality ranged from 1.9$^{\prime \prime}$ FWHM at $z$ to 2.1$^{\prime \prime}$ FWHM at $u$.

For VV 488 = MCG -02-58-11, we use the $B$ and $I$ images from the CTIO 1.5m telescope reported by \cite{WKC2000}.
 
It proved crucial to have these optical data at substantially higher resolution and signal-to-noise ratio than the
UV images. They show the context of each system and its level of symmetry (or departures from
symmetry) so that we can interpret the UV data with much greater confidence. 
The optical images were analyzed to yield maps of estimated attenuation at their full
resolution, or discover limiting factors which prevented doing so. 
In some cases,
despite a favorable geometry of the two galaxies, the optical data show no attenuation
in the apparent foreground galaxy (so either this is the background system or there is very
little dust along the backlit lines of sight). 
Some others are too
distorted to apply any of our symmetry approaches. Consequently, a comparatively small set of backlit
galaxies provides our information on the UV attenuation and its wavelength dependence. These categories of
pairs are listed in Table \ref{tbl-1} (for those with UV attenuation measures) and Table \ref{tbl-2} (not suitable
for such measures; some of these are still suitable for optical analysis).

\subsection {Analysis methods}
As set out by, for example, 
\cite{WK1992}, \cite{WKC2000}, and \cite{Holwerda2009},
the basic technique for retrieving attenuation from an overlapping galaxy system relies on the expression
$$ e^{- \tau} = {{ I - F} \over {B}} $$
where $I$ is the observed intensity at a given point, and $F$ and $B$ represent estimates of the intensities of foreground and background galaxy light
without any attenuation (using symmetry considerations). We often find it useful to work in transmission $T = e^{- \tau}$ rather than optical
depth $\tau$, since the errors from statistics of the data are better behaved and more symmetric than for $\tau$. We neglect effects of 
scattering; the relevant effect comes from the difference in light from the background galaxy scattered by the foreground between the
region under analysis and its symmetric point, which declines very steeply with the line-of-sight separation between the galaxies \citep{WKC2000}.
Depending on the structure of the galaxies and quality of the data available, we use one of three techniques exploiting different levels of symmetry
to estimate the foreground and background contributions to the light at each point.

Ellipse fitting: here we model one or both galaxies by fitting ellipses at small steps
in semimajor axis to the isophotes, then interpolating to make model two-dimensional images; we use the
{\it ellipse} and {\it bmodel} tasks within IRAF/STSDAS\footnote{IRAF is distributed by the National Optical Astronomy Observatory, which is operated by the Association of Universities for Research in Astronomy (AURA) under cooperative agreement with the National Science Foundation.
STSDAS is a product of the Space Telescope Science Institute, which is operated by AURA for NASA}, which implement the algorithm from
\cite{ellipse}. Errors in the estimates are are generated by propagating the scatter in values along the relevant isophotal ellipses for each galaxy,
propagating the standard deviation of these values through calculations of transmission and optical depth $\tau$.

Point symmetry: where the galaxies do not have the detailed symmetry needed for the 
other techniques, 
we use simple symmetry by rotating each galaxy image by 180$^\circ$, either for a full mapping of
the attenuation or measurement averaged over a resolved region. The error in modeling each galaxy in this way is evaluated
by the scatter in apertures of the same size located elsewhere on the relevant isophote for each galaxy

Arm tracing: here we assume that spiral arms have similar intensity profiles with radius, using
arms in a non-overlapped region as a guide to the behavior of ridge-line intensity with radius and the scatter 
due to fine-scale structure in the arms as evaluated at the relevant spatial resolution. 
Then the estimate of the non-backlit profile of an arm is taken by scaling the
reference arm profile to match non-backlit parts of the arm being analyzed, allowing intensity scaling. Our
technique for doing this starts by interactive marking of a number of points on the ridgeline of each
arm using a long-wavelength image ($I$ or $z$). These points are interpolated in polar coordinates
centered on the galaxy nucleus, sampled typically at 1$^\circ$ intervals. Then we can work with 
one-dimensional (resampled) profiles of intensity along each spiral arm as a function of position angle $\theta$.
The reference arm can be smoothed (typically using a median filter) to reduce the impact of discrete
star-forming regions on the profile shape. The error expected in using this as a model of the arm being
analyzed is taken from the scatter about the mean shape for multiple independent pieces of the
reference arm, of the same angular extent as the dust regions being measured. This approach reduces
our sensitivity to differences in the normalization of arm intensity or departures from exact $2 \theta$
symmetry in otherwise grand-design spirals.

Images were aligned using stars in common between optical and UV 
fields (or in a few sparse fields, using galaxy nuclei plus
scale and orientation parameters known from other fields).
We rebinned the UV data to the scale of the optical images
to avoid loss of information, and smoothed the optical data to the
resolution of the UV images with the appropriate Gaussian kernel
measured using stars in both image sets. Our general approach was to use the optical images, with better resolution and
signal-to-noise ratio, to guide our understanding of each pair's geometry, and then do
identical processing on registered and PSF-matched optical and UV images to
retrieve attenuation values for matching pieces of the galaxies.

To avoid potential biases in sample properties, we evaluated the suitability of each pair for these measurements
before measuring transmission. This was based based on geometry, surface brightness in the overlapping 
areas, and absence of UV-bright knots
in a position to distort measures in the overlap area. The regions analyzed were selected based on optimal location
for backlighting (combining background surface brightness with line of sight penetrating well into the foreground disk)
and evidence for dust from the higher-resolution optical images, to avoid contamination by (false) null detections from galaxies
which actually lie in the background.

\section{Extinction measures - individual systems}

\subsection{NGC 2207/IC 2163}

This pair is a very nearby interacting system. Extinction in the outer arms of NGC 2207 where they cross in front of
IC 2163 was addressed by \cite{Elmegreen2001}. Using the spatial resolution of HST data, they found diffuse
``intercloud" dust along the arms with $A_V = 0.5 -1.0$, and relatively discrete clouds with
$A_V=1-2$. The interaction between these galaxies has been modeled extensively. incorporating both 
morphological and kinematic information, with the most prominent distortion being the ocular form of IC 2163
\citep{Elmegreen1995}
This has, so far, left both galaxies symmetric enough to allow retrieval of dust attenuation in two arms
of NGC 2207, since the foreground arms are relatively dim where seen against bright parts of IC 2163.

\begin{figure*} 
\includegraphics[width=150.mm,angle=0]{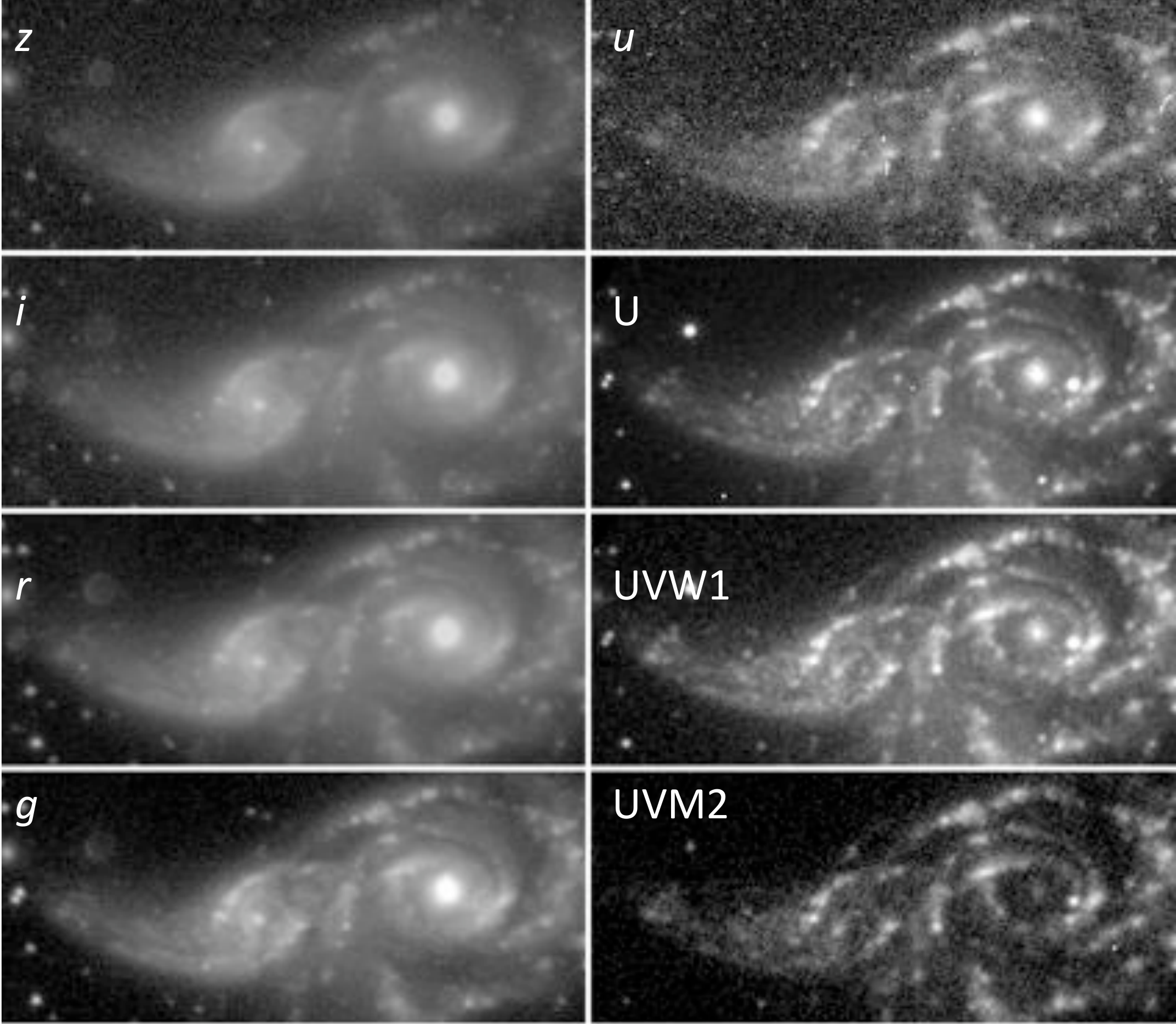} 
\caption{Montage of NGC 2207/IC 2163 as seen in $ugriz$ and XMM-OM UV bands.
Some bright stars have been patched by interpolation in the optical images.
The displays use an offset logarithmic intensity mapping, similar to the sinh
mapping recommended by \cite{color}, to retain detail over a wide 
dynamic range. The area shown spans
$254 \times 110$ arcseconds, with north at the top.} 
\label{fig-ngc2207}
\end{figure*}

For NGC 2207/IC 2163, we use the XMM-Newton Optical Monitor data in the UV, 
providing higher spatial resolution than GALEX. Fig. \ref{fig-ngc2207} compares
our images of this system from $z$ to UVM1 (0.95 - 0.25 $\mu$m).
Our interpretation is aided by archival HST WFPC2 images \citep{Elmegreen2000}.
The attenuation in NGC 2207 was measured most accurately using the arm-tracing
technique to estimate the background intensity and its error, and rotational symmetry for
the foreground arms in NGC 2207 itself; for the outermost arm in particular, the foreground intensity is so low that
its correction makes a negligible contribution to the error. 

\begin{figure*} 
\includegraphics[width=150.mm,angle=0]{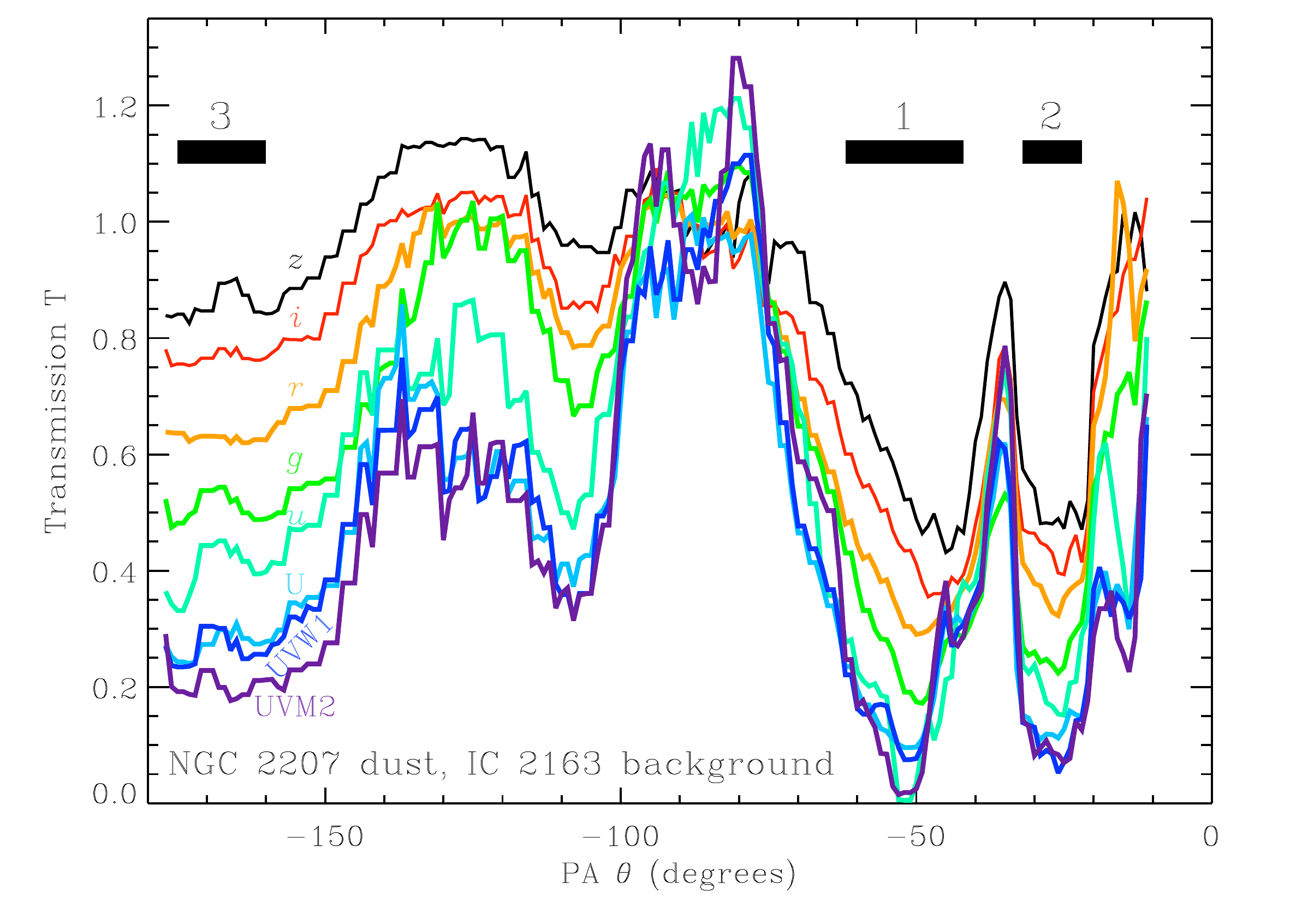} 
\caption{Slices of derived transmission $T$ along the background arm of IC 2163, as functions of 
position angle $\theta$. Lines for each filter band are coded in chromatic order, as marked by each curve
on the left. Errors are derived from the scatter in apparently absorption-free regions. 
The black boxes show the location and widths of three regions
defined as foreground arm segments for our analysis.} 
\label{fig-ngc2207arms}
\end{figure*}

Our most detailed results are shown in Fig. \ref{fig-ngc2207arms}, which compared the derived
transmission values along the background arm of IC 2163 as functions of position angle.
We trace the attenuation in three places where the outer arm of NGC 2207 crosses in front of the
western (inner) arm of IC 2163 to its NW (two distinct segments at PA --37$^\circ$) and one to the SW (PA -171$^\circ$).
The values averaged across these main foreground-arm crossings are listed with
our overall summary in Table \ref{tbl-3}. Errors are based on the scatter in regions with (nominally)
no foreground attenuation: $\theta = -95^\circ$ to -$70^\circ$ and $\theta= - 135^\circ$ to $-118^circ$.

From comparison with other systems, we also evaluated the multiband attenuation using
a single box location in the outer arm of NGC 2207, using simple symmetry to evaluate the
starlight contributions of both NGC 2207 and IC 2163 within this region. We use a $6 \times 17$-arcsecond
box with major axis oriented 23$^\circ$ clockwise from N, centered 3.5" W and 3.1" S of the 
nucleus of IC 2163.

The UV transmissions we measure suggest that the effective covering fraction of clouds with
significant optical depth must be much larger than in the optical. This must be so for a
realistic mix of pixel-by-pixel optical depths, with the result that the cross-section of spiral
features is larger than we see in higher-resolution optical images. We examine this for
NGC 2207 using the archival HST WFPC2 images described by \cite{Elmegreen2001}.
We compare distributions of retrieved transmission in the F336W and F439W filters,
approximately $U$ and $B$ bands respectively.

\begin{figure*} 
\includegraphics[width=125.mm,angle=0]{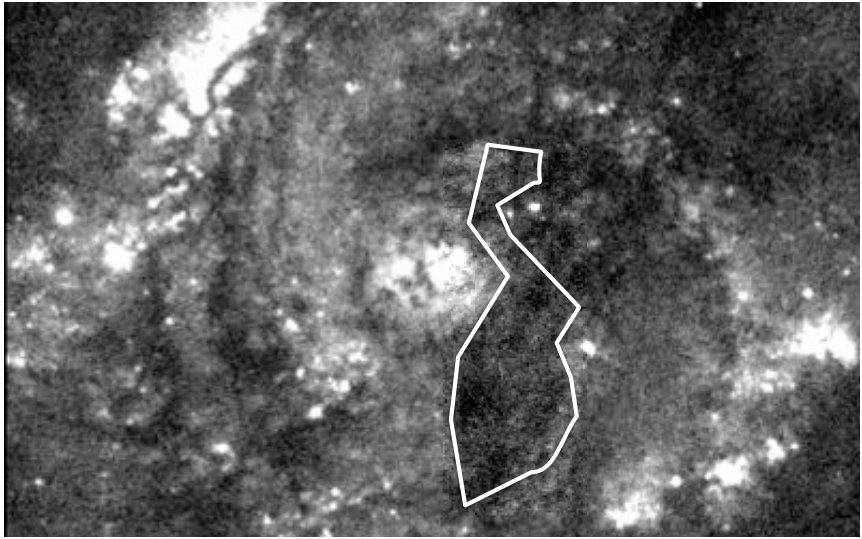} 
\caption{Region in the outer arm of NGC 2207 analyzed for Fig. \ref{fig-ngc2207hst}. The
HST WFPC2 F439W image is illustrated over a range $25.6 \times 40 ^{\prime \prime}$  around the center
of the background system IC 2163; the polygon encloses the pixel area modeled for
attenuation to compare covering fractions in F439W and F336W bands.} 
\label{fig-ngc2207region}
\end{figure*}

\begin{figure*} 
\includegraphics[width=125.mm,angle=0]{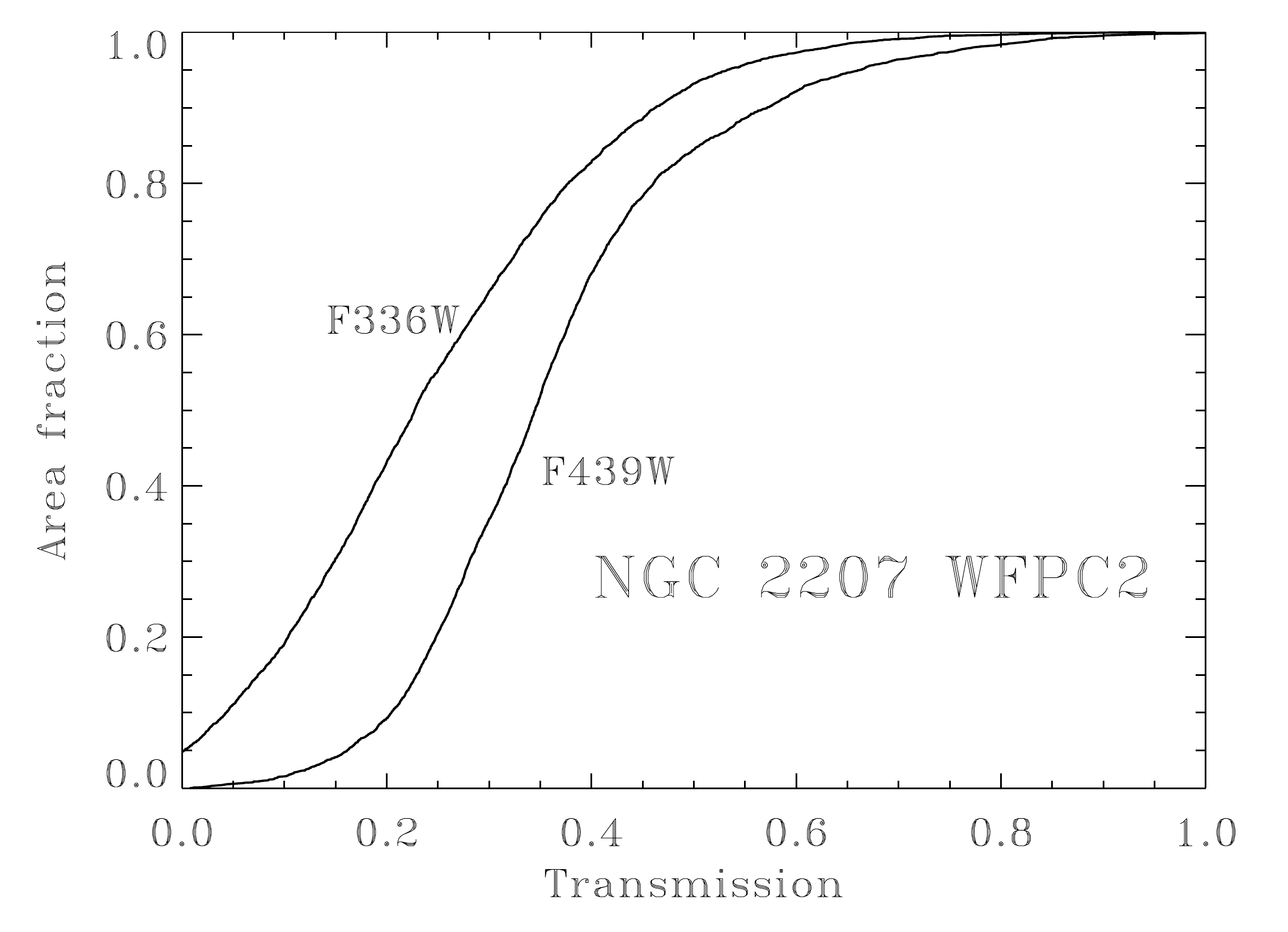} 
\caption{Distributions of transmission versus fractional area in the outer arm of NGC 2207
from HST images in the $U$ and $B$ bands; the region analyzed for this figure
is shown in Fig. \ref{fig-ngc2207region}. The comparison illustrates the increasing fraction of 
the arm area occupied by progressively greater optical depths toward shorter wavelengths.} 
\label{fig-ngc2207hst}
\end{figure*}

For the outer arm of NGC 2207 projected against the core of IC 2163 (Fig. \ref{fig-ngc2207region}), the foreground light correction 
is $< 20$\% in this region; we apply a constant surface brightness measured north of the bright disk
of IC 2163, and model the background light with a first-order radially symmetric distribution
(which is especially flat, essentially constant, in F336W). After smoothing over 0.5 $^{\prime \prime}$
due to the limited S/N in the $U$ filter, we derive distributions of relative area versus transmission within the dust lane (in a region
comparable to our UV measures) as shown in Fig. \ref{fig-ngc2207hst}.

\subsection{NGC 4567/8}

This spiral pair in the Virgo cluster, despite considerable overlap in our view, is not
strongly interacting. H I mapping, as shown by  \cite{Chung2009}, shows kinematic signs of disk
warping in the large foreground system NGC 4568; the two disk redshifts coincide quite closely in the overlap
region making it difficult to separate the two galaxies' contributions to the
H I column density in this region. Similarly, the CO velocity field in this region
shows a smooth connection, leading \cite{Kaneko} to assess the system as
in an early stage of interaction before first close approach.

\begin{figure*} 
\includegraphics[width=140.mm,angle=0]{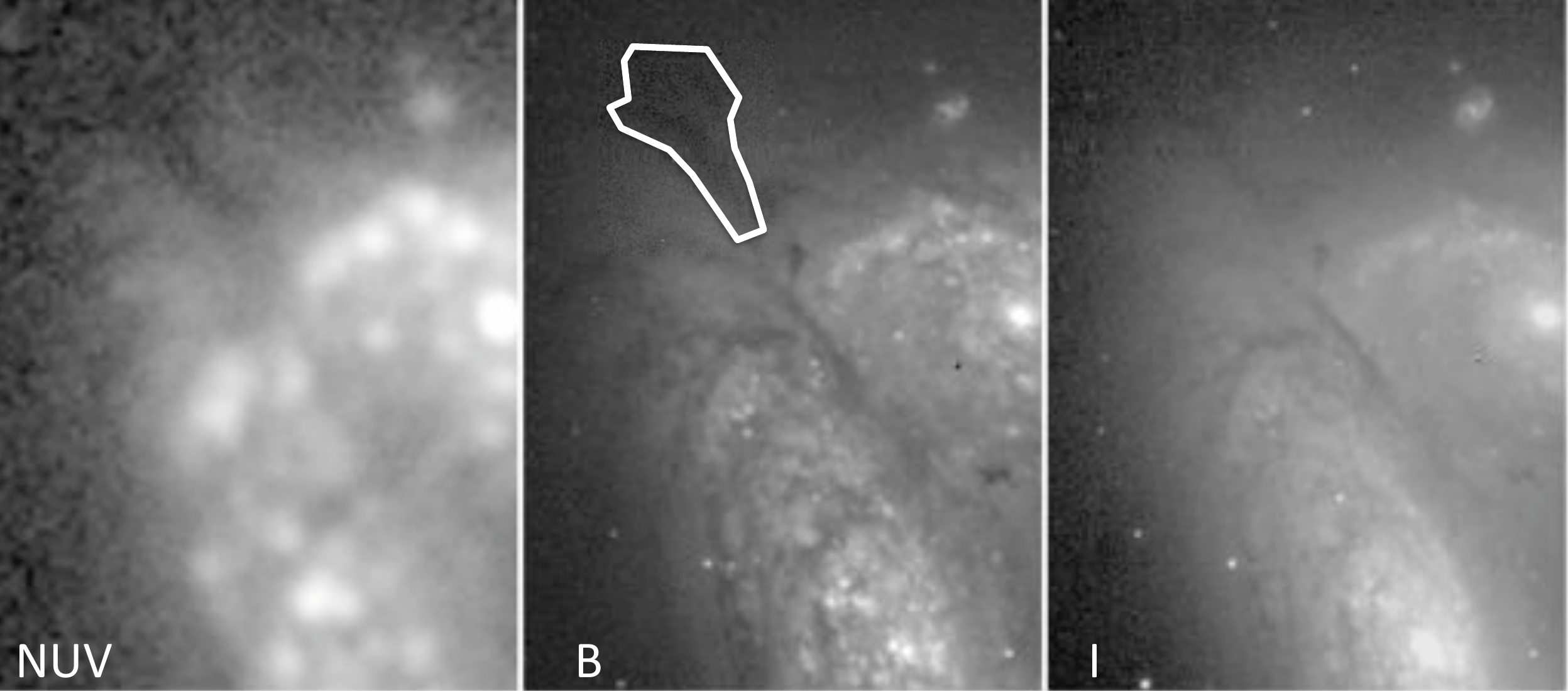} 
\caption{Overlap region of the pair NGC 4567/8 shown in GALEX NUV band
and optical $BI$ bands from WIYN OPTIC data. The dust lane analyzed here is indicated.
Images are display with an offset logarithmic intensity mapping, where brightness is proportional to
log (intensity + constant), approximating the
SDSS image rendering.
The area shown spans
$102 \times 36$ $^{\prime \prime}$, with north at the top.} 
\label{fig-ngc4568}
\end{figure*}

We measure a dust arm on the north side of NGC 4568, which lies far enough out in the
galaxy to minimize effects of stellar spiral structure on the surface brightness
(Fig. \ref{fig-ngc4568}). In this area we can use ellipse fits to the background galaxy NGC 4567
and assess errors from foreground structure from the scatter among similarly-sized regions on either side of the
dust lane after subtracting the background contribution. This dust feature, unlike other darker ones,
is clear of the luminous, UV-bright star-forming regions which are common in the inner disks
of both galaxies.

The transmission and derived attenuation values are listed in Table \ref{tbl-3}. The GALEX data give a usable measurement in the
NUV band; the available FUV exposure is too short for a detection at this surface brightness.

\subsection{UGC 3995}

This object, noted in the UGC \citep{UGC} as having a possible jet, was identified by  \cite{Keel1985} as a superimposed
galaxy pair. Both spirals are seen nearly face-on. Their matching redshifts indicate potential interaction, but the
spiral patterns are closely symmetric; potential distortions are at larger radii than we consider here.

``Snapshot" HST data in the F606W (approximately V) band by \cite{Malkan}
were analyzed initially by \cite{Marziani},
who quote  $A_V \approx ~0.18$ in the interarm parts 
of the foreground disk, and $A_V  >1.5 $ in the arms. Using the favorable
geometry, with the outer disk of UGC 3995B projected against the bulge of UGC 3995A,
\cite{HolKeel2013} have mapped the dust in more detail using the HST image
and integral-field spectra from the CALIFA program \citep{CALIFA},
noting a distinct outer edge to strong attenuation and a systematic offset between the arm
locations as defined by stars and dust.

The bulge component of UGC 3995A is faint in the UV, so we use arm-tracing rather
than ellipse-fitting for this system. To measure attenuation consistently with
wavelength, we use this approach for all wavelength bands in this work, rather
than adopting the symmetry-based results at high resolution from HST data by
\cite{HolKeel2013}.

As promising as the galaxy types look for UV imaging, the background arms are not
very blue, so there is not enough signal for measurement in the GALEX bands. We can
measure a region of the dust lane closest to the background nucleus in
all 5 SDSS filters including $u$, adding color information to the previous attenuation values
and, for this system, a bridge wavelength toward the average UV behavior.
At these wavelengths, the background light is so dominant that the corrections for foreground
light based on the opposite side of the disk are $<25$ \%, and the error contributions from these corrections
correspondingly smaller.

\begin{figure*} 
\includegraphics[width=140.mm,angle=0]{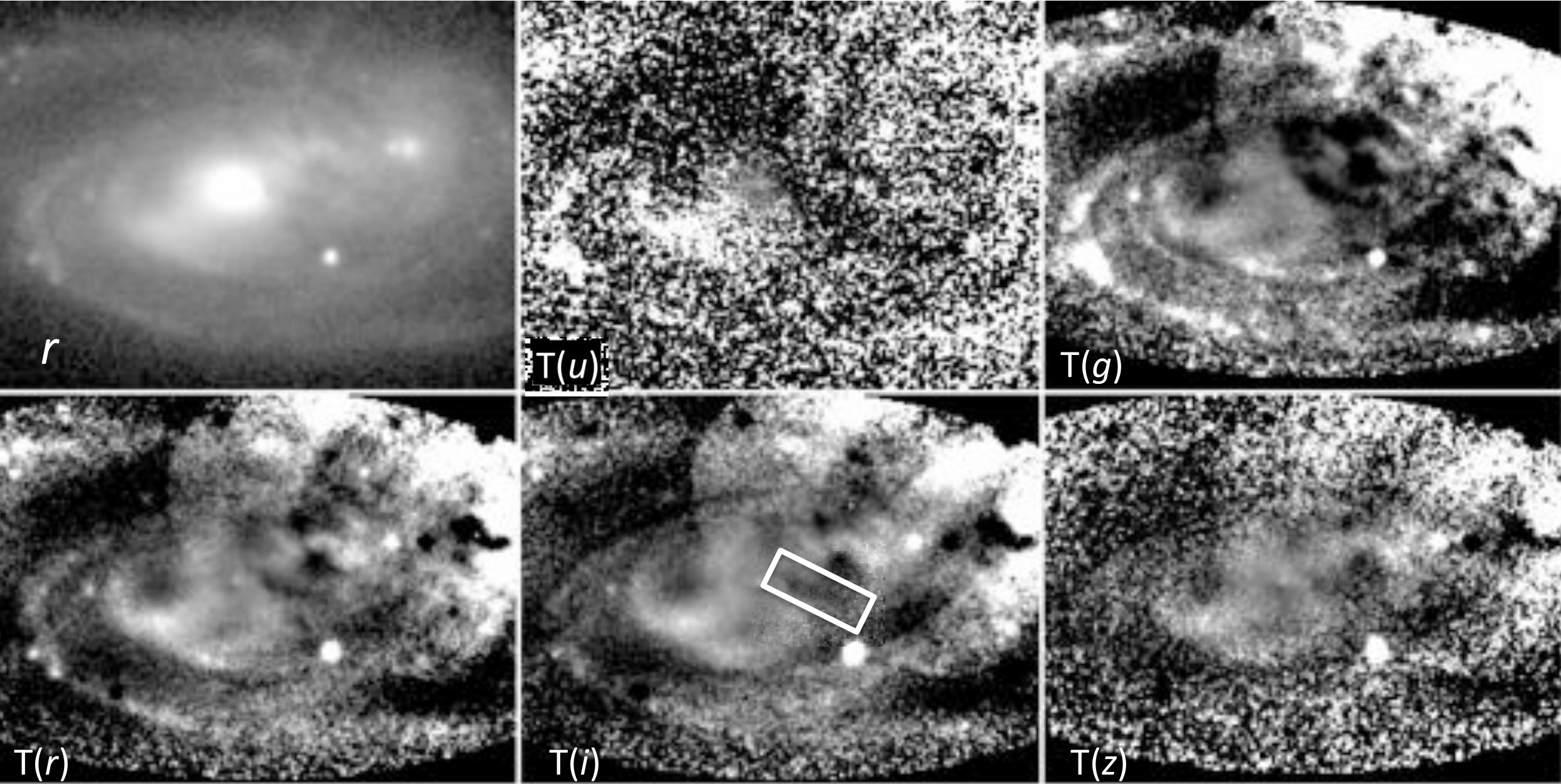} 
\caption{Maps of estimated transmission $T$ in SDSS $ugriz$ bands for
UGC 3995B. The strong variation in signal-to-noise is driven by the surface brightness
behavior of both foreground and background galaxies. The grey scale ranges from zero to 2 for the transmission maps, to 
sample the noise and symmetry artifacts properly. The $r$ image from the SDSS
is shown for comparison of the arm patterns. The box on the $i$ transmission map
shows the arm region used for attenuation measurements here. The region shown spans $ 85 \times 63  ^{\prime \prime}$. }
\label{fig-ugc3995}
\end{figure*}

Our transmission maps (Fig. \ref{fig-ugc3995}) show the effect noted by \cite{HolKeel2013} in which
the dust arms lie inward of the bright stellar ridge lines, typically by $\approx 1.5 ^{\prime \prime}$ or 500 pc.

\subsection{SDSS J143650.57+060821.4}
We trace intensities along arms in both systems. This pair (hereinafter SDSS 1436 for brevity) is favorable in that the arms of the background, face-on
galaxy are bluer than those of the foreground galaxy, so they are brighter in the UV away from the foreground dust and
the error contributed by foreground structure is minimized. We have both WIYN OPTIC (BI) and KPNO 2.1m (BR) 
optical images, with slightly better surface-brighness sensitivity for the 2.1m data. The sense
of overlap is clear from the light loss against the eastern arm of the western galaxy where the eastern
arm of the other galaxy crosses it; the more face-on and symmetric galaxy is in the background.

\begin{figure*} 
\includegraphics[width=140.mm,angle=0]{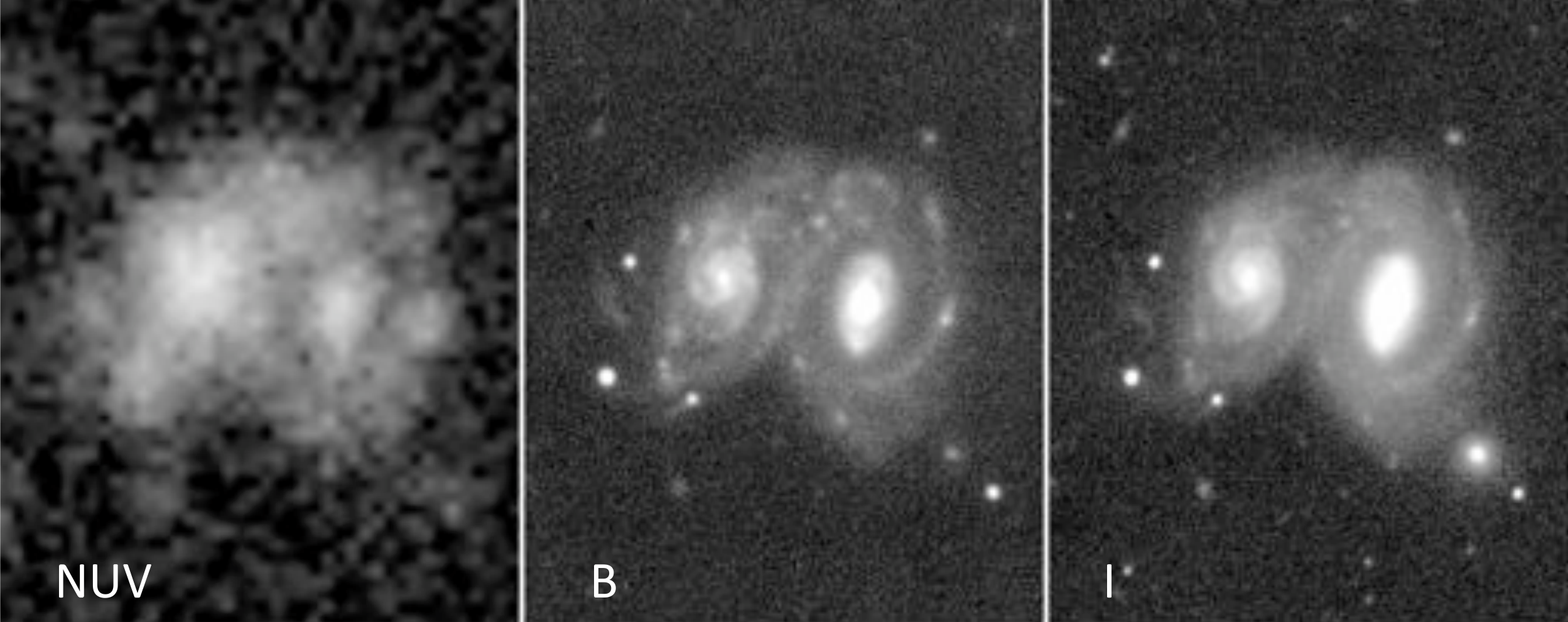} 
\caption{Galaxy pair SDSS J143650.57+060821.4 shown in GALEX NUV band
and optical $BI$ bands from WIYN OPTIC data. 
Images are display with an offset logarithmic intensity mapping, approximating the
SDSS image rendering.
The area shown spans
$60 \times 73$ $^{\prime \prime}$, with north at the top.} 
\label{fig-sdss1436}
\end{figure*}

These arms are nearly tangent to each other in projection; we measure transmission in a region where they
are closest (Fig. \ref{fig-sdss1436}). Errors are based on the scatter among similarly-sized slices of each arm (compared to its symmetric partner)
in non-overlapped regions. Even if we consider the arm profiles ignoring the UV-bright knot opposite
the overlap region, the NUV transmission has an error range spanning all physical values.

\subsection{SDSS J161453.42+562408.9}
\begin{figure*} 
\includegraphics[width=140.mm,angle=0]{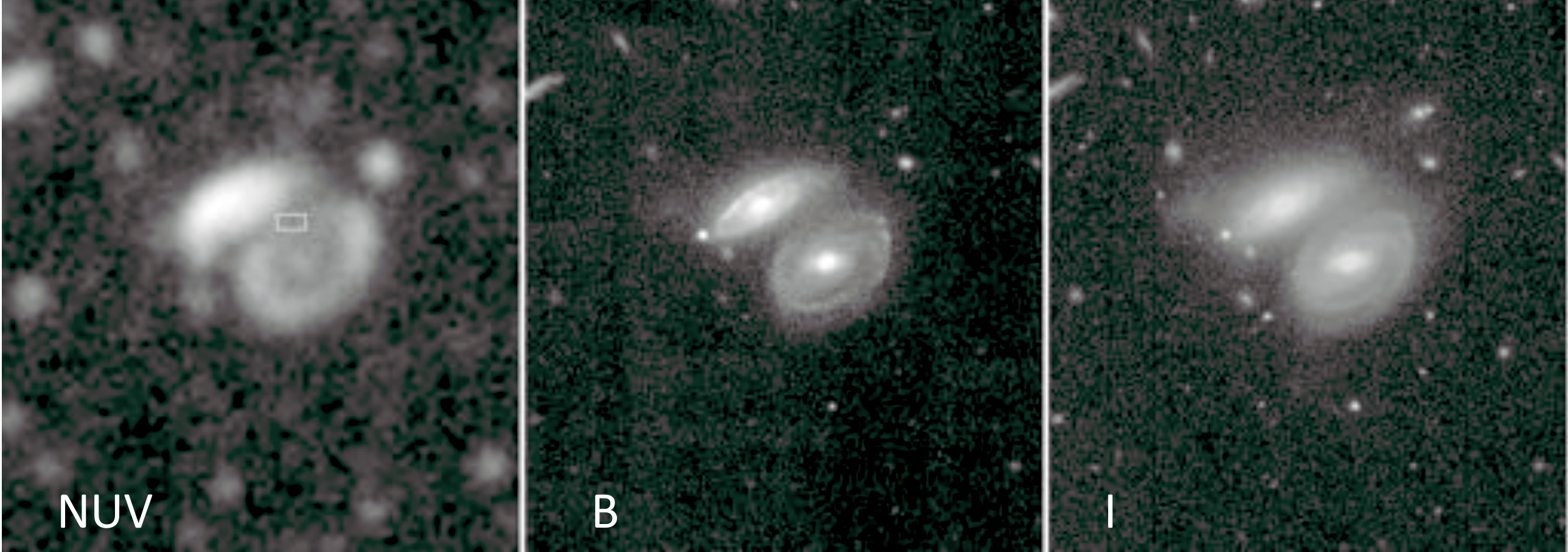} 
\caption{Images of SDSS J161453.42+562408.9 with GALEX in NUV band, and with the WIYN OPTIC system in B and I. 
Offset logarithmic intensity mapping was use for each to span a wide dynamic range.
The box indicates the region in which attenuation was measured. The region shown spans $ 106 \times  115 ^{\prime \prime}$
with north at the top. }
\label{fig-sdss1614}
\end{figure*}

The evidence for attenuation of light from the southern member of this pair by the northern galaxy is a pronounced
deficit in the NUV intensity, with a much weaker $B$ deficit in the same area (Fig. \ref{fig-sdss1614}). We estimate 
the foreground and background contributions in the overlap region by reflection symmetry, and evaluate errors
from the scatter of multiple regions of the same size along the relevant global isophotes for each galaxy.

\subsection{SDSS J163321.48+502420.5}   

\begin{figure*} 
\includegraphics[width=140.mm,angle=0]{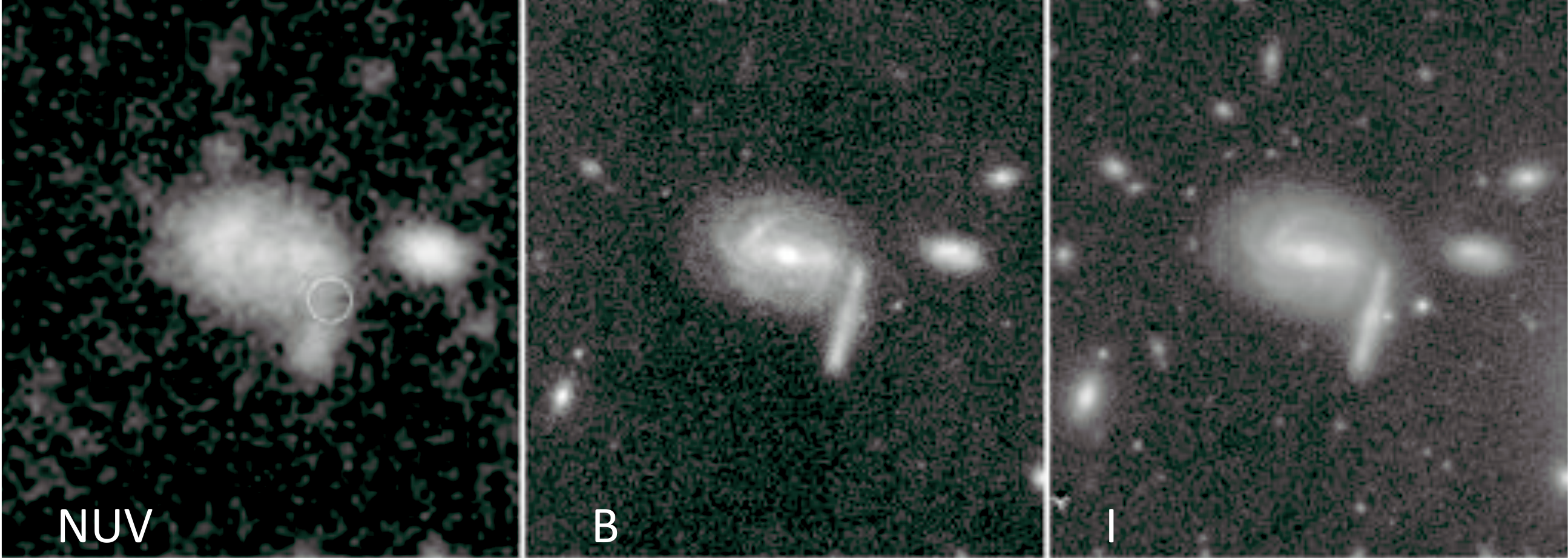} 
\caption{Images of SDSS J163321.48+502420.5 with GALEX in NUV band, and with the WIYN OPTIC system in B and I. 
Offset logarithmic intensity mapping was use for each to span a wide dynamic range.
The circle indicates the region in which attenuation was measured. The region shown spans $ 106 \times  115 ^{\prime \prime}$
with north at the top.}
\label{fig-sdss1633}
\end{figure*}

The attenuation in this region appears in front of the northern part of the edge-on background galaxy, We use symmetry for both
galaxies, modeling the NUV image of the foreground galaxy
with a linear decline in surface brightness (a close fit which allows easy evaluation of scatter about the mean behavior).
In this system, we measure the attenuation in a single box region (Fig. \ref{fig-sdss1633}), where the error is dominated by foreground structure
and evaluated by the scatter along the overlap isophote elsewhere in the foreground system.

\subsection{SDSS J211644.67+001022.4}

\begin{figure*} 
\includegraphics[width=140.mm,angle=0]{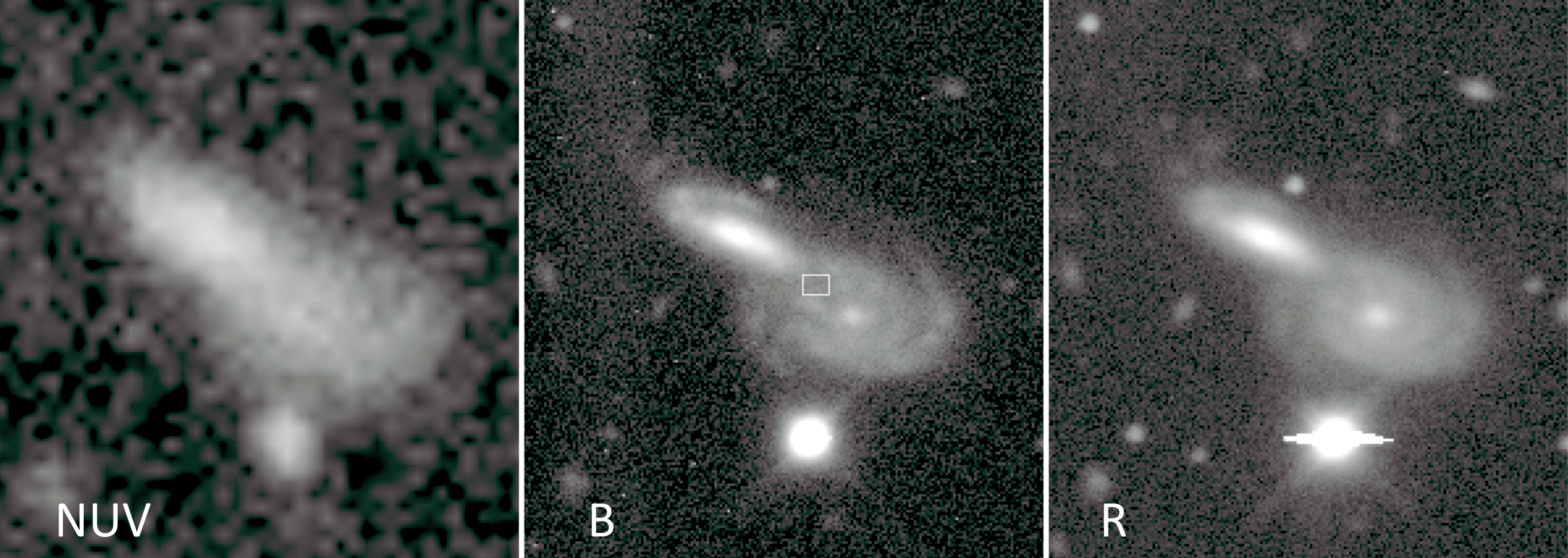} 
\caption{Images of SDSS J211644.67+001022.4 with GALEX in NUV band, and with the KPNO 2.1m imager in B and R. 
Offset logarithmic intensity mapping was use for each to span a wide dynamic range.
The box indicates the region in which attenuation was measured. The region shown spans $ 60 \times   65^{\prime \prime}$
with north at the top.}
\label{fig-sdss2116}
\end{figure*}

We use simple reflection symmetry to model the galaxy light, and evaluate the errors based on scatter of
multiple regions of the same size along the relevant isophote of each galaxy. The system geometry and dust
region are shown in Fig. \ref{fig-sdss2116}.

\subsection{NGC 5491}

\begin{figure*} 
\includegraphics[width=140.mm,angle=0]{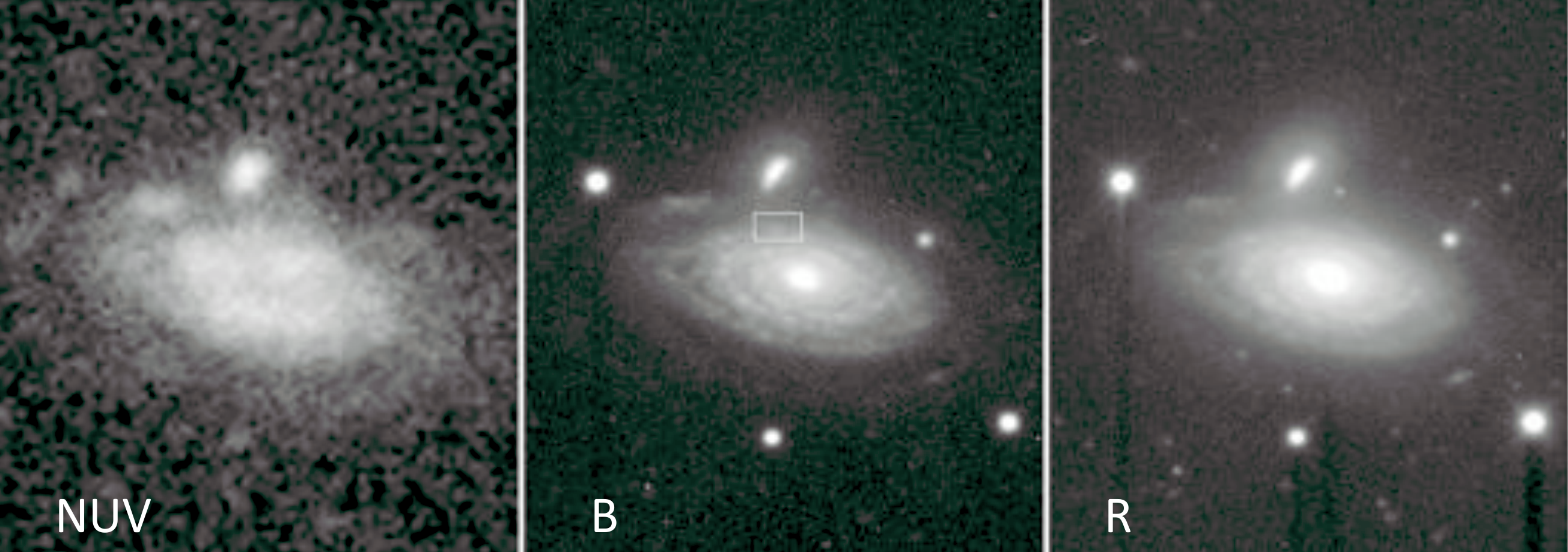} 
\caption{Images of NGC 5491 with GALEX in NUV band, and with the KPNO 2.1m in B and R. 
Offset logarithmic intensity mapping was use for each to span a wide dynamic range.
The box indicates the region in which attenuation was measured. The region shown spans $ 121 \times   130^{\prime \prime}$
with north at the top.}
\label{fig-ngc5491}
\end{figure*}

The evidence for attenuation is somewhat ambiguous in this pair, consisting of the drop in surface brightness of the outer arm of
the southern spiral where it is projected against the (apparently) foreground northern companion. Using reflection symmetry,
and evaluating errors from scatter in same-sized regions along the relevant isophotes in each galaxy, our data suggest actual
light loss, clearly detected in both GALEX UV bands but only marginally in the optical bands. The geometry and region measured
are shown in Fig. \ref{fig-ngc5491}.

\section{Attenuation measures - average behavior and reddening law}

We combine the results from these galaxy pairs to constrain the mean attenuation law, connecting light loss and reddening. In doing so,
we consider the error ranges as they affect not only the weighting of data, but normalization to the
optical values. The optimum attenuation range in this respect has the $B$ values large enough to
be well-measured, and the UV values not so large as to be nearly indeterminate. The combined
values are listed in Table \ref{tbl-4}.

\begin{figure*} 
\includegraphics[width=125.mm,angle=0]{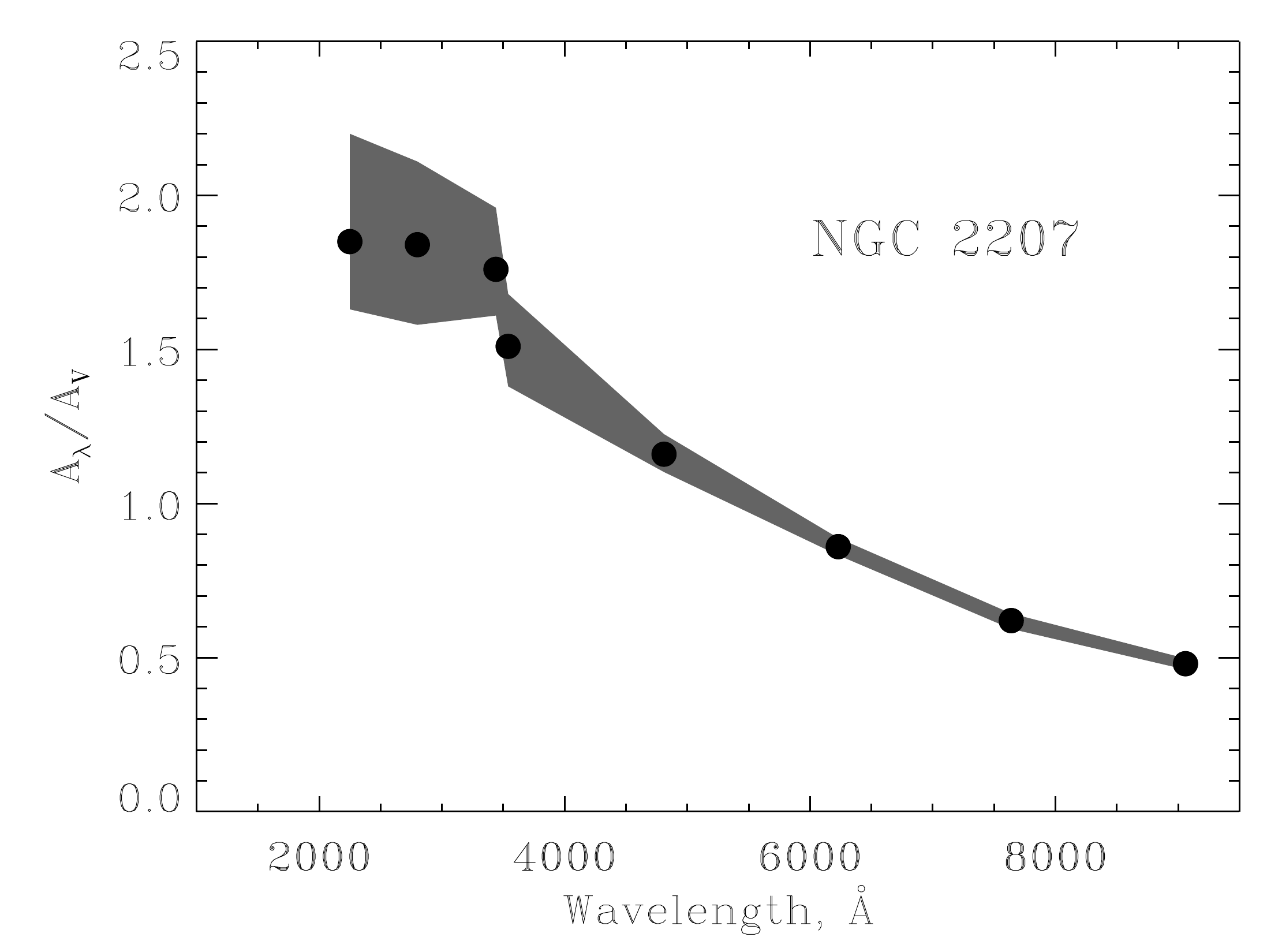} 
\caption{Wavelength dependence of attenuation in NGC 2207. The normalization simply interpolates
$A_V$  between the values for $g$ and $r$. Points show variance-weighted means,
and the shaded region shows the area spanned by the standard deviation of these weighted means.}
\label{fig-ngc2207extplot}
\end{figure*}

Since the data for NGC 2207 have high quality for each region, we show its values separately
(Fig. \ref{fig-ngc2207extplot}). The shaded region encompasses the errors on
variance-weighted means in each filter, normalized to $V$ by interpolation between $g$ and $r$ bands.

For the systems with GALEX images, we can extend the attenuation data to the FUV band near 1500 \AA. The large range in
S/N of these measures dictates care in combining data for the various galaxies. Some systems have
the optical points measured very poorly, so they would compromise the normalization of the UV values.
Accordingly, we normalize the NUV/B ratio via the weighted mean of objects with the B values measured
at S/N$>1$ (in practice, this means S/N$>2.5$). Likewise, we normalize the composite FUV/NUV ratio with 
a weighted mean of the 2 systems with non-open error bars (limits for other systems are not informative).

\begin{figure*} 
\includegraphics[width=125.mm,angle=0]{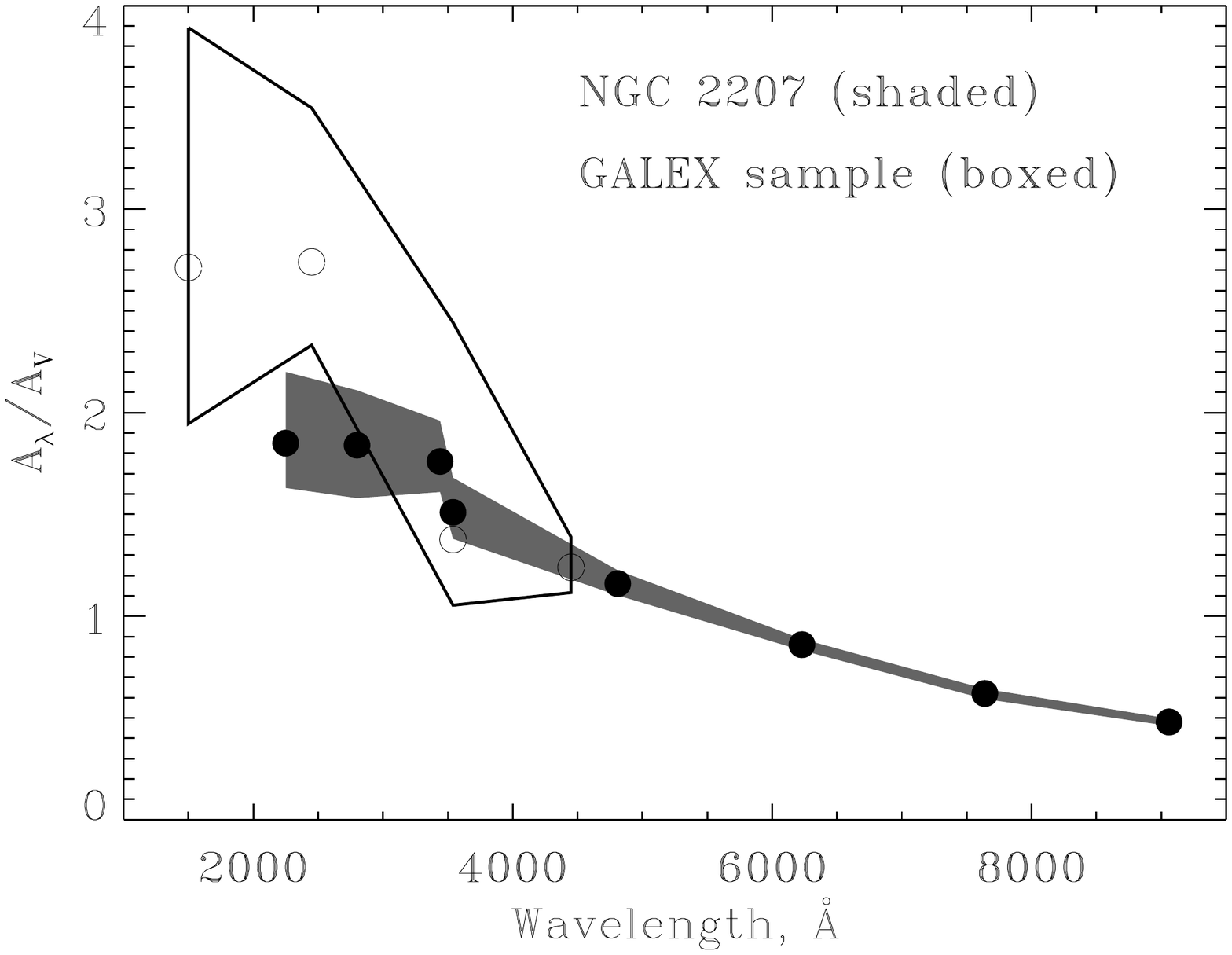} 
\caption{Wavelength dependence of attenuation in the GALEX subsample, compared to NGC 2207. 
Combination of data at various wavelengths and S/N levels is described in the text; the $u$band
point comes solely from UGC 3995B.
The normalization in the optical simply assumes
that $A_B$ matches the behavior in NGC 2207. Open circles show variance-weighted means,
and the boxed region shows the area spanned by the standard deviation of these weighted means.
The error bars are noticeably asymmetric in the UV bands, being symmetric in transmitted
intensity so that the range extends farther to higher attenuation.}
\label{fig-extsummary}
\end{figure*}

Fig. \ref{fig-extsummary} compares these results to the NGC 2207 means. While the averages are somewhat
higher than for NGC 2207, the error bounds (at the standard deviation level) overlap almost everywhere, so that at our
precision a single attenuation law is a plausible fit to the whole data set. The error band is broad
enough for such subtle effects as shifts of effective wavelength with reddening not to be of concern.

\section{Radial behavior in the ultraviolet}

We can examine the radial behavior of the attenuation, mindful of the inherent selection caused by the
need to have transmitted light in the UV bands. Therefore,
our most accurate results occur for transmission values not very close to either zero or unity,
where the relative effects of symmetry errors are greatest. 
Arm and interarm regions are not clearly distinguished at the resolution of  the GALEX data;  only in SDSS 2116 is it likely that
the UV measure is clearly dominated by arm dust. Taking the entire sample, the arm regions do show systematically
greater attenuation at a given normalized radius $R/R_{25}$ than combined arm/interarm values,
but our errors are too large to compare attenuation laws in spiral arms and between them. The high arm 
value are dominated by NGC 2207, for which the XMM-OM UV data do resolve the arms well.

\begin{figure*} 
\includegraphics[width=125.mm,angle=0]{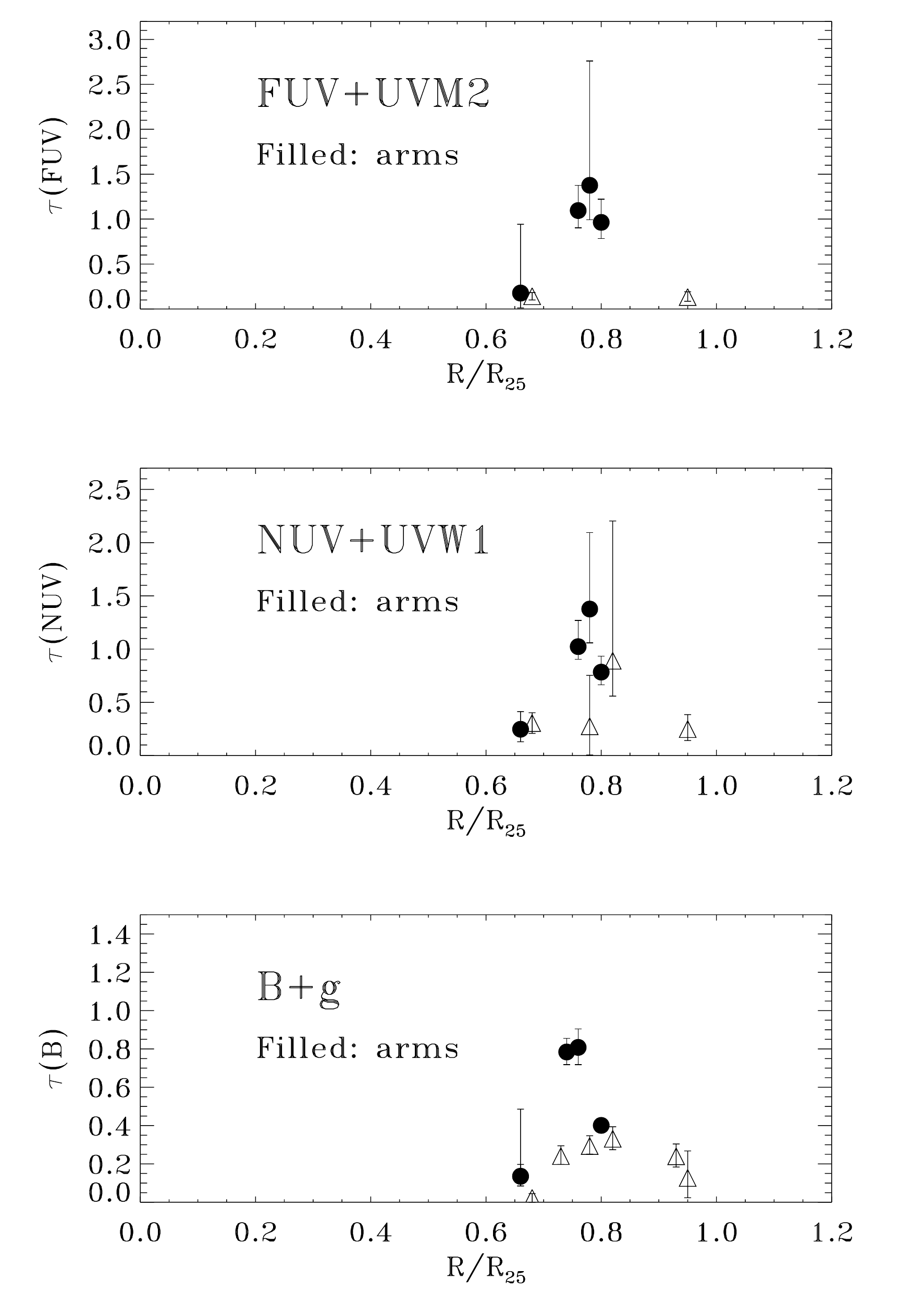} 
\caption{Derived optical depth $\tau$ as a function of normalized radius $R/R_{25}$. Adjacent bands are plotted together, where the systematic changes
are less than the errors in points. Filled circles indicate points identified
as being dominated by dust in individual spiral arms, in NGC 2207 and SDSS 2116. Open triangles indicate
measurements averaged over larger regions in all other systems.
Two points in the $B/g$ plot have been shifted by 0.02 in $R/R_{25}$ to avoid overlap.}
\label{fig-radial}
\end{figure*}

Within these limitations, we summarize in Fig. \ref{fig-radial} the UV attenuation as a function
of normalized radius, corrected to equivalent face-on values with a cosine factor (for the assumed
thin-disk geometry, the same as axial ratio $b/a$). The major (unsurprising) feature of these plots is that 
data identified as arm-dominated show higher attenuation than the other points averaged across arm and
interarm regions. 

\cite{Boissier} discuss the radial behavior of attenuation in the spiral M83 = NGC 5236 from a combination of UV, H$\alpha$ and infrared
tracers. These will likely be weighted toward star-forming regions, and might be expected to give a higher value than
our area-weighted data. The outer edge of their relation, at $R/R_{25}=1.0$, has A$_{FUV} = 1.1-1.7$ magnitude.
For our sample, the derived face-on values are 0.6 -- 1.9 over the range $R/R_{25} = 0.66-0.95$. The 
UVM2 filter data at a similar wavelength  for the arms of NGC 2207 also fit broadly with their results, $A = 1.0-1.5$ at $\approx 0.8 R_{25}$.

Optical dust measurements from our data can naturally be performed and interpreted at much higher resolution than their
continuation into the UV; we will present such results for a large sample of backlit spiral elsewhere, including both grand-design
and flocculent systems.

\section{Conclusions}

We have used a combination of ground-based optical images with GALEX and XMM-OM data in the
ultraviolet to measure the shape of the effective extinction (attenuation) law in galaxy disks over the range 1500-9000 \AA\ .
Only spiral background galaxies are bright enough at the UV wavelengths to serve for this technique, so
the structure in both galaxies contributes significantly to the
error in retrieving the transmission fractions, and subsequently optical depths. The data for the nearby
spiral NGC 2207 are particularly good; since it was also observed with different filters than the GALEX system, we
consider its attenuation behavior separately. 

The effective extinction law derived by \cite{calzetti} from the SEDs of star-forming galaxies has found wide
applicability. It pertains to large areas of galaxies, and, like our results, shows a flatter slope than the intrinsic grain behavior
seen in star-by-star investigation of nearby resolved galaxies (\citealt{ccm} , \citealt{Bianchi}, \citealt{smcdust}). \cite{calzetti} attribute this to a mix 
of effects: the progressive bias in favor of the more transparent parts of a patchy dust distribution at shorter 
wavelengths, and the potential effects of preferential escape of longer-lived stars from obscuring
clouds around regions of recent star formation.

\begin{figure*} 
\includegraphics[width=125.mm,angle=0]{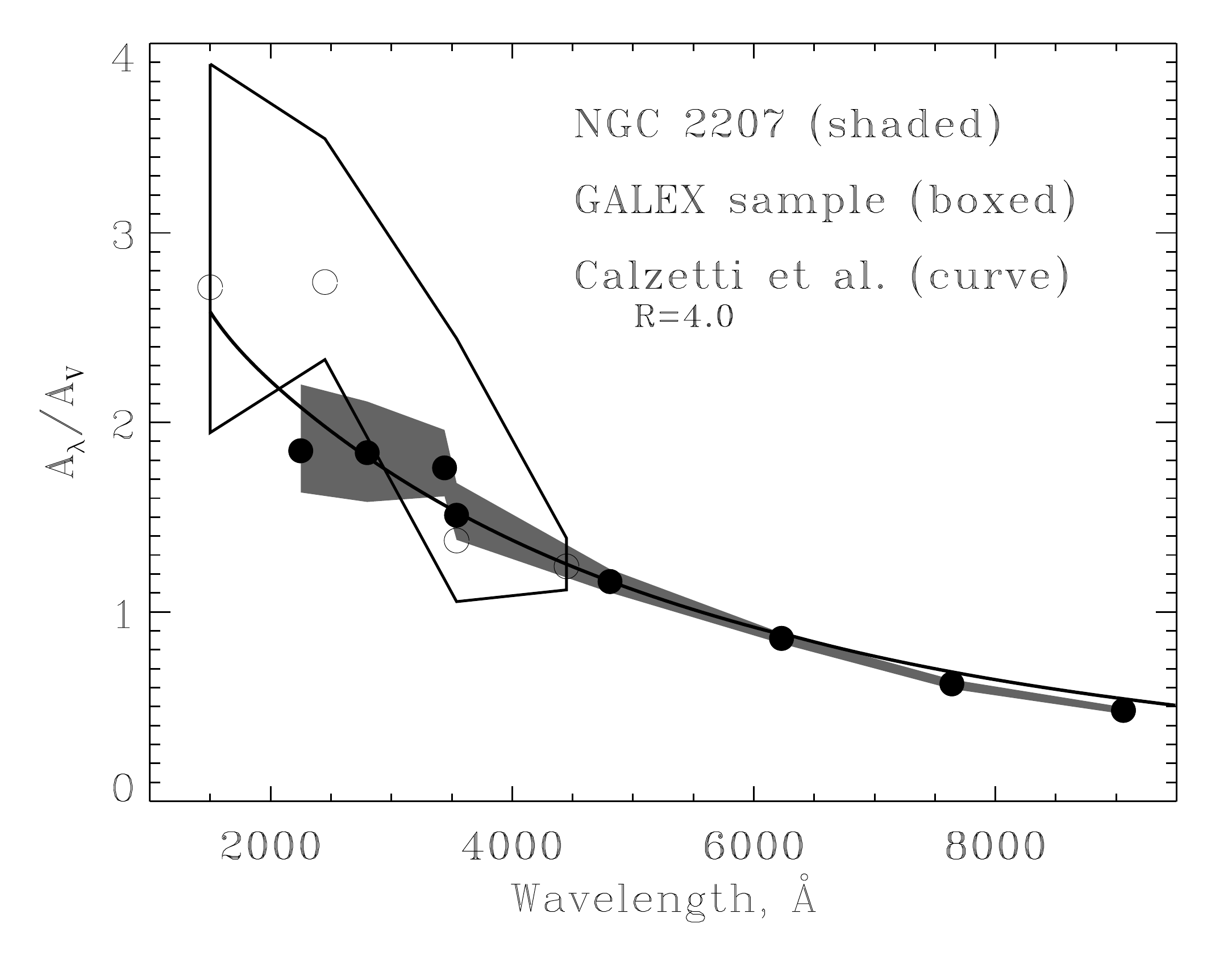} 
\caption{Comparison of our mean data, as in Fig. \ref{fig-extsummary}, to the Calzetti et al. form,
using their analytic expression. The assumed value of $R_V$ affects both normalization and slope, since
our technique can measure grey attenuation as well as reddening. The best-fit value assuming no grey
component is $A_V=4.0$, close to their value of 4.06 from SED fitting.}
\label{fig-calzetti}
\end{figure*}

The \cite{calzetti} law is a remarkably good fit to our observations (Fig. \ref{fig-calzetti}). SED fitting is insensitive to 
a grey component of attenuation; if we assume there is no actually grey offset, only a single free parameter
remains affecting both normalization and slope, equivalent to $R_V = A_V / E_{B-V}$. The curve in Fig. \ref{fig-calzetti}
shows the best-fit value for our data, $R_V = 4.0 \pm 0.1$, closely comparable to their derived mean $R_V=4.06$.
Our results extend the applicability of the \cite{calzetti} form, to even the outer parts of spiral disks where star formation
proceeds at very modest levels.

Particularly in the case of NGC 2207, the foreground-light correction and its error are small enough to insure that
the slope of the reddening curve we find is not strongly affected by differential escape of stars from obscuring clouds;
fine structure in the dust distribution must account for the attenuation behavior we find. In nearby, well-resolved spirals
analyzed using background galaxies, the reddening slope changes with spatial resolution in a way consistent
with a fractal cloud distribution on scales from tens to hundreds of parsecs \citep{KW2001a}. This raises at least
the possibility that different dust distributions could occur, giving different reddening behavior, which suggests
caution in applying these results to galaxies at high redshifts.

\begin{figure*} 
\includegraphics[width=125.mm,angle=0]{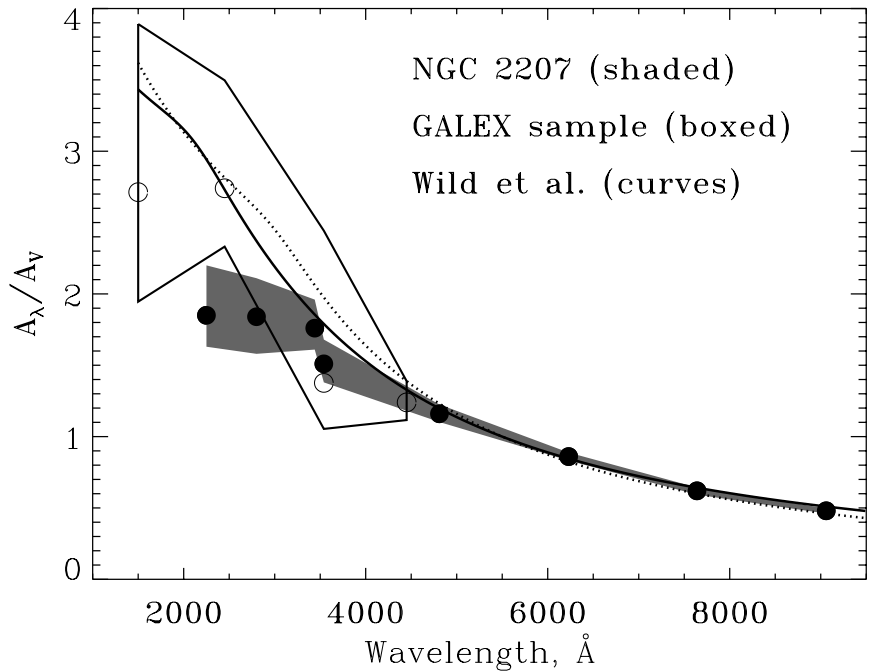} 
\caption{Comparison of our mean data, as in Fig. \ref{fig-extsummary}, to Wild et al.,
using their analytic expression. Their expression includes stellar mass, axial ratio, and specific
star-formation rate (SSFR) as parameters. We show predicted attenuation relations for the mean axial ratio $b/a=0.61$
of our sample, evaluated at extremes of $3 \times 10^{10}$ solar masses and SSFR$= 10^{-9}$ year$^{-1}$, as a
solid curve, and
$3 \times 10^{8}$ solar masses and SSFR$= 10^{-7}$ year$^{-1}$ , plotted as a dashed curve. These are both consistent with
the mean for our GALEX sample, but not with the UV results for NGC 2207}
\label{fig-wild}
\end{figure*}

\cite{Wild} performed an SED analysis using many pairs of galaxies, matched in specific star-formation rate, metallicity,
and axial ratio, to form 
attenuation curves for subgroups of galaxies, expressing the results as spliced power-law segments varying with each
of the matching variables. For their results
as well, comparison with our mean attenuation data may help separate the roles of dust distribution itself from the
relative distributions of dust and stars. Our composite curve is consistent with their UV slopes, steeper than the
Calzetti et al. value, over wide
ranges in stellar mass and star-formation rate, for the mean of the GALEX sample, but not for NGC 2207 (Fig. \ref{fig-wild}). Using
their expressions for more inclined disks reduces but does not eliminate this discrepancy; our sample has a mean
axial ratio $ < b/a> = 0.61$, with range 0.32-1.0 (Table \ref{tbl-3}). As \cite{Wild} note, much of the difference they find with
axial ratio and star-formation rate may trace to distinct populations of grains near star-forming regions and in the
diffuse ISM; the attenuation statistics with area in NGC 2207 suggest that a proportionally greater fraction  of the UV extinction
arises in more diffuse material.

Taken together, our data suggest a flatter UV attenuation curve than any of the Wild et al. forms; this could reflect the nature of
backlighting measurements, which are area-weighted independent of the location of stars in the foreground galaxy. The
GALEX sample by itself is reasonably well fit by the Wild et al. curve, however; the errors for these points are substantially
larger than for the NGC 2207 data, leaving open the possibility of a difference.

\begin{figure*} 
\includegraphics[width=125.mm,angle=0]{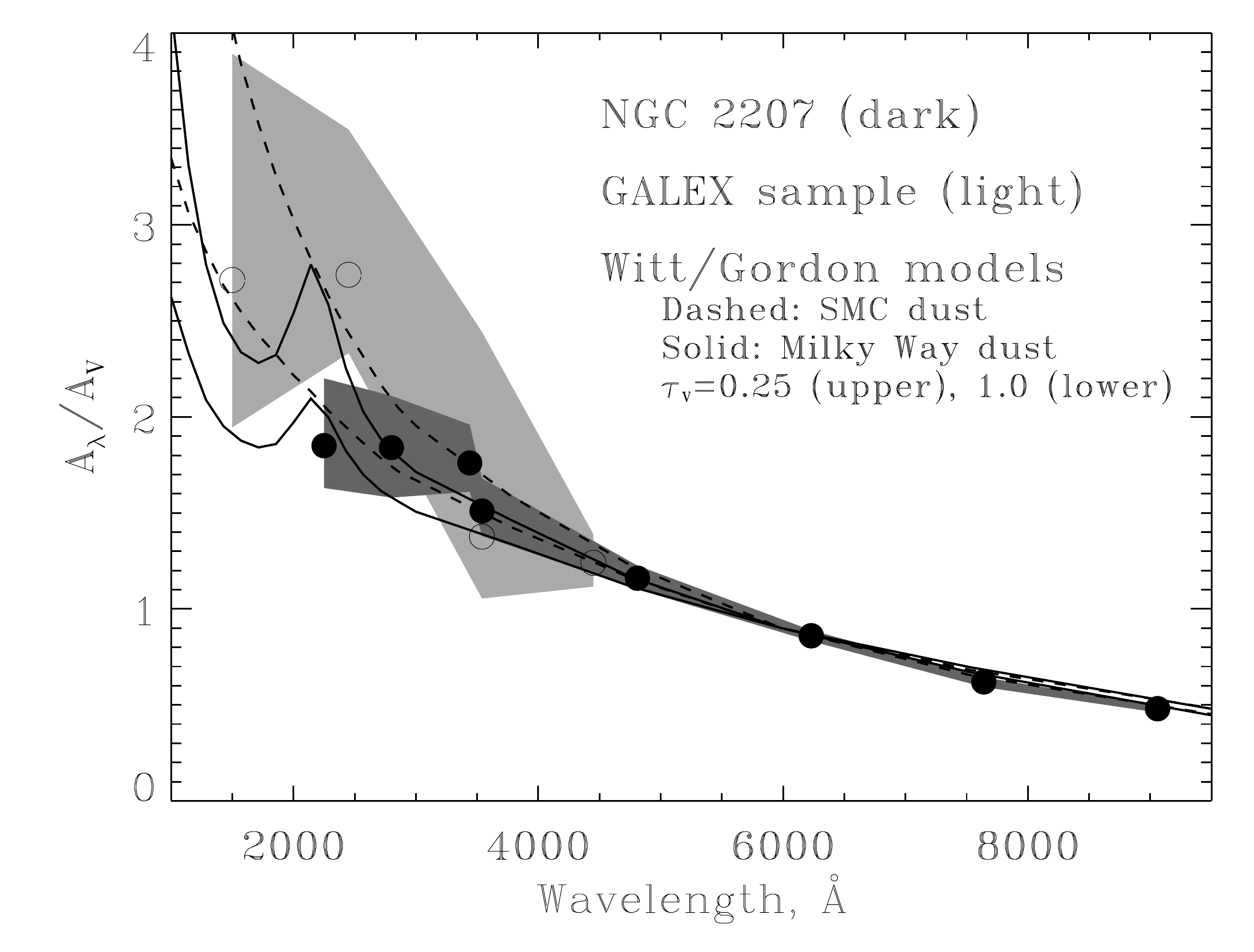} 
\caption{Comparison of our mean data, as in Fig. \ref{fig-extsummary}, to the calculations from
\cite{WittGordon2000} for attenuation by both Milky Way and SMC-like grains, for various degrees of dust
clumping. For each grain type, the upper curve is for averaged $\tau_V=0.25$, and the lower one for $\tau_V=1$. 
For the entire UV data set, either $\tau_V=1$ curve is acceptable in a $\chi^2$ sense, with the NGC 2207 results
providing a stronger preference for the same models, implying a strong role for clumping. For clarity, shaded regions denote $\pm 1 \sigma$ error bounds
for each data subsample.}
\label{fig-models}
\end{figure*}

This situation clearly represents a sheet geometry, reasonably well approximating the shell geometry used in 
some radiative-transfer calculations. Calculations have been published including realistic degrees of clumping which
we can compare to our results, for the cases of Milky Way and SMC-like dust populations (e.g. \citealt{WittGordon2000}).
For this situation, internal scattering effects will be removed by symmetry, so the more relevant comparison is with
the ``direct" component of attenuation excluding scattering. The \cite{WittGordon2000} results, provided
in tabular detail,  include multiple levels of scaling
to optical depth averaged over all lines of sight; for a fixed attenuation along one line of sight; this amounts to changing the
degree of clumpiness of the dust. We show a comparison in Fig. \ref{fig-models}. Using a simple $\chi^2$ figure of
merit, the entire UV data set ($\lambda < 3000$ \AA ) favors both $\tau_V=1$ calculations (with $\chi^2 \approx 1.3$) 
over the $\tau_V=0.25$ values ($\chi^2 \approx 2$); 
that is, the scenarios with clumpier grains. For the more precise NGC 2207 data, the evidence is even stronger, $\chi^2 \approx 0.5$ versus 
$\chi^2 \approx 3$. The comparison is somewhat degenerate even between such extremes as Milky Way and SMC dust first because
the NUV and UVM2 bands overlap with the 2175-AA extinction feature, and because increased clumping manifests itself in the calculations
as a flatter UV slope in each case, progressively masking the difference due to grain extinction properties.

Our results suggest the applicability of something very close to the \cite{calzetti} form for effective extinction of kpc-scale
areas in the outer disks of spirals. This requires a mild extrapolation to the equivalent optical value, indicating that
further work can use this technique to compare dust content and distribution of galaxies at significant redshifts observed in the emitted UV to 
the present-day galaxy population as understood in the optical regime.

\acknowledgments

This project was enabled by many of volunteer participants in the Galaxy Zoo project;
their contributions to the backlit-galaxy program are acknowledged individually at http://data.galaxyzoo.org/overlaps.html.
This work was supported by the NASA Astrophysics Data Program (ADP) under grant NNX10AD54G.
Some of the data presented in this paper were obtained from the Mikulski Archive for Space Telescopes (MAST). 
STScI is operated by the Association of Universities for Research in Astronomy, Inc., under NASA contract NAS5-26555. 
Support for MAST for non-HST data is provided by the NASA Office of Space Science via grant NNX09AF08G and by 
other grants and contracts. This research has made use of NASA's Astrophysics Data System, and
the NASA/IPAC Extragalactic Database (NED) which is operated by the Jet Propulsion Laboratory, California Institute of Technology, 
under contract with the National Aeronautics and Space Administration.

C. J. Lintott acknowledges funding from The Leverhulme Trust and the STFC Science in Society Program. K.
Schawinski was supported by the Henry Skynner Junior Research Fellowship at Balliol College,
Oxford and by a NASA Einstein Fellowship at Yale, and gratefully acknowledges support
from Swiss National Science Foundation Grant PP00P2 138979/1. Galaxy Zoo was made
possible by funding from a Jim Gray Research Fund from Microsoft and The Leverhulme Trust.

Funding for the creation and distribution of the SDSS Archive has been provided by the Alfred
P. Sloan Foundation, the Participating Institutions, the National Aeronautics and Space Administration,
the National Science Foundation, the U.S. Department of Energy, the Japanese Monbukagakusho,
and the Max Planck Society. The SDSS Web site is http://www.sdss.org/. The
SDSS is managed by the Astrophysical Research Consortium (ARC) for the Participating Institutions.
The Participating Institutions are The University of Chicago, Fermilab, the Institute for Advanced
Study, the Japan Participation Group, The Johns Hopkins University, Los Alamos National
Laboratory, the Max-Planck-Institute for Astronomy (MPIA), the Max-Planck-Institute for Astrophysics
(MPA), New Mexico State University, Princeton University, the United States Naval Observatory,
and the University of Washington.




\clearpage







\clearpage

\begin{table}
\begin{center}
\caption{Galaxy pairs analyzed.\label{tbl-1}}
\begin{tabular}{lccccccc}

\tableline\tableline
System & $z_{\rm fg}$  &  R$_{25}$ & \multispan2 GALEX exp (s):  & Optical data 
  &  \multispan2 Models: \\
&   & arcsec & NUV & FUV &  & Foreground & Background \\
\tableline
NGC 2207	& 0.0091 & 127 & 8919&  13380       &       SARA-S        & arm tracing &  arm tracing \\
NGC 4568	& 0.0075 & 137 & 10170 & 1706 	&	HST, WIYN &	ellipse & ellipse\\
NGC 5491         & 0.0197 & 14 & 1967 & 1322            &  KP2m           & symmetry   &  symmetry\\
UGC 3995	&  0.0158 & 27 & 1535 & 1535		& HST, SDSS       & symmetry  &  ellipse\\
SDSS J143650.57+060821.4	& 0.0588 & 19 & 2485 & 2485		& KP2m            & arm tracing & arm tracing\\
SDSS J161453.42+562408.9       & -- & 20 & 21858 & 6118          &   WIYN,KP2m   &    symmetry  &  symmetry\\
SDSS J163321.48+502420.5	& 0.0439  & 21 & 2698 & 2698		& WIYN          &  ellipse & mirror symmetry\\
SDSS J211644.67+001022.4	&  0.0318 & 16 & 4513 & 4513		& KP2m       &     symmetry  &  symmetry\\
\tableline
\end{tabular}
\end{center}
\end{table}


\clearpage

\begin{table}
\begin{center}
\caption{Unsuitable galaxy pairs for UV analysis.\label{tbl-2}}
\begin{tabular}{lcccl}
\tableline\tableline
{}  & \multispan2 GALEX exp (s): & \\
System &   NUV & FUV & Optical data & Problem \\
\tableline
NGC 4231  &       11734  & 11734       &      WIYN      &      too far apart\\
NGC 4911    &      1688 & 1688            &   WIYN, HST &  bg too dim in UV\\
NGC 5679      &     1670 & 1670           &   HST,WIYN     &   UV  resolution too poor\\
NGC 6365   &   961     &             0           & KP2m       &     symmetry failure in UV\\
SDSS J084726.06+533814.9   &    2922 & 1696             &   WIYN    &  too dim at overlap\\
SDSS J102517.76+170821.0   &    1722 & 1722             & WIYN      &      too far apart\\
SDSS J103244.35+543847.5   &    1680 & 1680           &   WIYN      &      symmetry problem \\
SDSS J105454.32+100250.0 &      1551 & 1551             & WIYN       &       bg too small/faint in UV\\
SDSS J121326.98+504237.4 &     2440 & 1288            &   WIYN        &    UV surface brightness too low\\
SDSS J121626.29+470131.6 &    11774 & 11774         &    WIYN       &     interacting, distorted\\
SDSS J124415.63+314242.6  &   3370 & 1672            &  WIYN          &  interacting, distorted\\
SDSS J125725.25+272416.4  &    31099 & 29932         &     WIYN       &    fg asymmetry\\
SDSS J131222.90+461906.1  &    3116 & 1584            &  WIYN         &   warped, bad symmetry, faint UV\\
SDSS J131354.20+441048.6  &    1442 & 1592            &  WIYN         &   bg too dim in UV  \\
SDSS J131404.57+472145.3    &    2540 & 1655             &  WIYN        &   no absorption detected\\
SDSS J142718.87-014042.4  &    1680 & 1680            &  WIYN         &   too far apart - arms in wrong place\\
SDSS J154954.44+085140.6   &    1424 & 1058            &   WIYN        &    bg too dim in UV\\
UGC 6212    &  	 3213 & 3213               & 	WIYN         &   overlap surface brightness too low\\
VV 488       &   629   &  0	&	CTIO 1.5m     &     background  too dim in UV\\
\tableline
\end{tabular}
\end{center}
\end{table}

\begin{deluxetable}{lcccccccc}
\tabletypesize{\scriptsize}
\tablecaption{Summary of Attenuation Results \label{tbl-3}}
\tablewidth{0pt}
\tablehead{\colhead{Foreground galaxy}     & \colhead{Type}  &  \colhead{$b/a$} &  
\colhead{PA$^\circ$} & 
\colhead{R/R$_{25}$} & \colhead{Band} & \colhead{Transmission}&  \colhead {$\tau$}   }

\startdata
NGC 2207 (box) & SAB(rs)bc	 & 0.65	& 100	 & 	0.78 & UVM2  & $0.50 \pm  0.66$ & - \\ 
			 &          		&		&		&	& UVW1 & $0.63\pm 0.36$ & $0.46^{+0.85}_{-0.45}$\\
			 &          		&		&		&		& OM U & $0.64\pm 0.27$ & $0.45^{+0.54}_{-0.36}$ \\
			 &          		&		&		&		& $u$   &  $0.57 \pm 0.28$ &   $0.56^{+0.68} _{-0.40}$\\
			 &          		&		&		&		& $g$   &   $ 0.61 \pm 0.05$ & $0.49^{+0.09}_{-0.07}$ \\
			 &          		&		&		&		& $r$     & $ 0.67 \pm 0.05$ & $0.40^{+0.09}_{-0.07}$ \\
			 &          		&		&		&		& $i$	   & $0.74 \pm 0.08$ & $0.30^{+-.11}_{-0.10}$ \\
			 &          		&		&		&		& $z$    & $ 0.78 \pm 0.15$ &  $0.25^{+0.21}_{-0.18}$ \\
NGC 2207 region 1	&		&		&		&	0.76 & UVM2 &  $0.16 \pm 0.06$ & $1.83^{+0.47}_{-0.32}$ \\
			 &          		&		&		&		& UVM1	& $0.18 \pm 0.06$ & $ 1.71^{+0.41 }_{-0.20 }$ \\
			 &          		&		&		&		& OM U	& $0.18 \pm 0.05$ & $ 1.71^{+0.33 }_{-0.24 }$ \\
			 &          		&		&		&		& $u$ &	$0.17 \pm 0.05$ & $ 1.77^{+0.35 }_{-0.26 }$\\
			 &          		&		&		&		& $g$ & $ 0.27 \pm 0.03$ & $ 1.31^{+ 0.12}_{-0.11 }$\\
			 &          		&		&		&		& $r$ & $0.36 \pm 0.02$ & $ 1.02^{+0.06 }_{-0.05 }$\\
			 &          		&		&		&		& $i$	 & $0.46 \pm 0.02$ & $ 0.78^{+0.04 }_{-0.04 }$\\
			 &          		&		&		&		& $z$ & $0.56 \pm 0.02$ & $ 0.58^{+0.04 }_{-0.04 }$\\
NGC 2207 region 2	&		&		&		&	0.78 & UVM2	& $0.10 \pm 0.09$ & $2.30 ^{+2.31}_{-0.64 }$\\
			 &          		&		&		&		& UVM1	& $0.10 \pm 0.07$& $ 2.30^{+1.20}_{-0.53 }$ \\
			 &          		&		&		&		& OM U & $ 0.13 \pm 0.06$ & $ 2.04^{+0.62}_{-0.38}$\\
			 &          		&		&		&		& $u$ & $	0.20 \pm 0.07$ & $ 1.61^{+0.43}_{-0.30}$\\
			 &          		&		&		&		& $g$ & $	0.26 \pm 0.04$& $ 1.35^{+0.16}_{-0.15}$ \\
			 &          		&		&		&		& $r$ &$0.37 \pm 0.03$ & $ 0.99^{+0.09}_{-0.07}$\\
			 &          		&		&		&		& $i$	& $0.43 \pm 0.02$ & $ 0.84^{+0.05}_{-0.04}$\\
			 &          		&		&		&		& $z$ & $	0.50 \pm 0.02$& $ 0.69^{+0.04}_{-0.04}$ \\
NGC 2207 region 3	&		&		&		&	0.80 & UVM2 & $0.20 \pm 0.07$ & $ 1.61^{+0.43}_{-0.30}$\\
			 &          		&		&		&		& UVM1 & $0.27 \pm 0.06$ & $ 1.31^{+0.25}_{-0.20}$\\
			 &          		&		&		&		& OM U	& $0.27 \pm 0.05$& $ 1.31^{+0.20}_{-0.17}$ \\
			 &          		&		&		&		& $u$ & $0.41 \pm 0.06$ & $ 0.89^{+0.16}_{-0.13}$\\
			 &          		&		&		&		& $g$ & $0.51 \pm 0.03$ & $ 0.67^{+0.06}_{-0.05}$\\
			 &          		&		&		&		& $r$ & $0.63 \pm 0.02$ & $ 0.46^{+0.03}_{-0.03}$\\
			 &          		&		&		&		& $i$	 & $0.77 \pm 0.02$ & $ 0.26^{+0.03}_{-0.02}$\\
			 &          		&		&		&		& $z$ & $0.86 \pm 0.02$& $ 0.15^{+0.02}_{-0.02}$ \\
NGC 4568	&	Sbc	         & 	0.39	& 27	         &   0.73	& NUV & $0.16 \pm	0.15$ & $>1.18$ \\
			&			&		&		&		& B	 & $0.51 \pm 0.07$ & $0.67^{+0.15}_{-0.12}$ \\ 
			&			&		&		&		& I	& $0.92 \pm  0.04$ &  $0.08^{+0.05}_{-0.04}$ \\ 
NGC 5491	& 	SBc:	 	 & 0.81	& 137       & 0.68         &   FUV & $ 0 .62 \pm 0.09$ & $ 0.48^{+0.14}_{-0.14}$ \\ 
			&			&		&		&		& NUV  & $ 0.66 \pm 0.08$ &  $0.41^{+0.13}_{-0.11}$ \\ 
			&			&		&		&		& B &     $1.00 \pm 0.03 $ &  $ <0.03 $ \\    
			&			&		&		&		& R     & $ 1.03 \pm 0.03$ &  $<0.03$ \\    
UGC 3995		& Sbc	& 1.0		& 0.89 & 0.95	&   $u$ & $0.70 \pm 	0.08$ & $0.36^{+0.12}_{-0.11}$ \\ 
			&			&		&		&		& $g$ & $	0.77	\pm 0.07$ & $0.26^{+0.07}_{-0.06}$ \\ 
			&			&		&		&		& $r$ & $0.84	\pm 0.03$ & $0.18^{+0.03}_{-0.03}$ \\ 
			&			&		&		&		& $i$  & $0.85	\pm 0.03$ & $0.16^{+0.04}_{-0.04}$ \\ 
			&			&		&		&		& $z$ & $0.93 	\pm 0.05$ & $0.07^{+0.06}_{-0.04}$ \\ 
SDSS J143650.57+060821.4 & Sc &    0.58 & 154 &   0.75 &    NUV & $<1.09$ & - \\
			&			&		&		&		& B	& $1.02 \pm  0.10$ & $<0.12$ \\
			&			&		&		&		&R   & $0.89 \rm 0.05$ & $ 0.12 ^{+0.05}_{-0.06}$ \\ 
SDSS J161453.42+562408.9		& Sbc &  	0.51	& 119 & 0.95 & FUV & $0.64 \pm 	0.12$ & $0.45^{+0.20}_{-0.16}$ \\ 
			&			&		&		&		& NUV & 	$0.59  \pm 0.15$ &$	0.53^{+029}_{-0.23}$ \\  
			&			&		&		&		& B	& $ 0.76  \pm 0.20$ &	$0.27^{+0.30}_{-0.22}$ \\ 
			&			&		&		&		&  	I &  $0.84  \pm	0.22$  & $ <0.47$ \\ 
SDSS J163321.48+502420.5 & SBbc     &         0.62  &  75 & 0.82 & NUV &  $0.21  \pm 0.19$ & $1.56^{+2.3}_{-0.58}$ \\ 
			&			&		&		&		& B	& $0.56  \pm 0.06$ &   $ 0.58^{+0.11}_{-0.10}$ \\ 
			&			&		&		&		& I	& $0.76  \pm 0.05$ &   $  0.27^{+0.07}_{-0.06}$ \\ 
SDSS 211644.67+001022.4	& Sc	  & 	0.32&  59	& 0.66      &  FUV  &$0.55\pm 0.51$ &  $ 0.60^{+2.6}_{-0.56}$ \\ 
			&			&		&		&		& NUV & $	0.43   \pm  0.18$ & 	$0.84^{+0.55}_{-0.40}$ \\
			&			&		&		&		& B	& $0.63 \pm 0.12$ &  $0.46^{+0.21}_{-0.17}$ \\ 
			&			&		&		&		& R &  $0.72 \pm 0.07$ &	$0.33^{+0.10}_{-0.12}$ \\ 
\enddata
\end{deluxetable}

\begin{table}
\begin{center}
\caption{Weighted mean attenuation behavior\label{tbl-4}}
\begin{tabular}{lcc}
\tableline\tableline
Sample & Band & $A_\lambda / A_V$ \\
\tableline
NGC 2207 &  UVM2 &   $1.85^{ +0.35}_{ -0.22}$ \\
 		 & UVW1 & $1.84^{+0.27}_{-0.26}$\\
		& U        & $   1.76^{+0.20}_{-0.15}$ \\
		& $u$      & $1.51^{+0.17}_{-0.13}$ \\
		& $g$      &  $1.16^{+0.065}_{-0.058}$\\
		& $r$       & $ 0.86^{+0.028}_{-0.028}$\\
		& $i$      & $ 0.62^{+0.023}_{-0.027}$\\
		& $z$     & $ 0.48^{+0.017}_{-0.019}$\\
GALEX& FUV & $2.71^{+1.12}_{-0.87}$ \\ 
    		& NUV & $2.74^{+0.75}_{-0.40}$ \\
		& $u$ & $1.38^{+0.60}_{-0.53}$ \\
		& B & $1.24^{+0.12}_{-0.10}$ \\
\tableline
\end{tabular}
\end{center}
\end{table}



\end{document}